# Elasto-hydrodynamics of droplet-pool-interactions

**Md Sultan**[#] and **Purbarun Dhar**[*]

Hydrodynamics and Thermal Multiphysics Lab (HTML), Indian Institute of Technology Kharagpur, West Bengal–721302, India

[#]E–mail: mdsultan1489@gmail.com

**Corresponding author*: E–mail: purbarun@mech.iitkgp.ac.in

## Abstract

In Newtonian fluids, impact of a droplet on a liquid pool births a cavity, crown, capillary waves, and Worthington jet. The corresponding hydrodynamic events for elastic (or Boger) fluids, however, remain an uncharted domain of comprehension and exploration. We thoroughly investigate, via experiments, theory, and simulations; how elastic energy storage, fluid relaxation, and competitive inertio-elasto-capillarity govern the spatio-temporal evolution of the cavity, the crown, and the ensuing Worthington jet in polymeric elastic fluids. The events are systematically explored over a wide range of impact Weber ($We$) and Deborah ($De$) numbers, considering varied Newtonian and elastic fluid droplet-pool combinations; and revealing new, and distinct morphological regimes compared to Newtonian counterparts. We illustrate that these new findings are purely driven by fluid-elasticity, and not by viscosity or interfacial tension. We derive a theory for cavity radius evolution, using energy conservation within potential-flow framework. We show that ~30-40 % of the droplet's kinetic impact-energy may be stored as elastic energy by the stretching polymer chains during cavity expansion. Appealing to the FENE-P model, we derive a theory for the temporal evolution of the radius of the elongated Worthington jet. We show that in elasto-capillary regime, competitive elastic- and capillary- stresses lead to exponential decay of the jet radius. The role of elastic stresses and the local velocity field in governing cavity evolution, morphology, and jet formation are further elucidated through computer simulations. Our findings significantly advance the uncharted paradigm of interplay between inertia, capillarity, and elasticity in droplet-pool interaction elasto-hydrodynamics.

## 1. Introduction

Hydrodynamic events consequent to the impact of a droplet on the free surface of deep, liquid pools have been extensively studied; albeit restricted to purely Newtonian fluid systems (droplet and pool liquids). Worthington pioneered in exploring the fluid dynamic events associated with the impact of a liquid droplet against a target body of liquid (Worthington 1908), which subsequently attracted academic researchers down the century towards understanding the phenomena. Droplet impact on liquid surfaces may be commonly encountered in engineering systems, such as ink-jet printing, spray cooling, and spray painting and coating (Yarin 2006). Concurrently, uncontrolled and ballistic development of such Worthington jets result in aerosol formation from ocean surfaces, soil erosion, and pesticide wastage and over-usage (Joung & Buie 2015; Nuruzzaman *et al.* 2016).

Upon droplet impact on a deep liquid pool, the cavity formed expands freely to attain a maximum size, thereby displacing a substantial volume of liquid in the form of a rising crown. At higher impact velocity, the crown may disintegrate into a large number of tiny droplets. The cavity expansion is only temporary, as surface tension gradually counteracts, and eventually arrests the downward motion of the liquid. Following this arrest, the flow reverses and the cavity undergoes collapse. The merging of capillary waves ejects a Worthington jet from the cavity's centre. This high-speed jet then undergoes Plateau-breakup, resulting in the formation of one or more droplets travelling upwards (Castillo-Orozco *et al*. 2015). During cavity retraction, different sections of the cavity wall may merge, leading to the entrapment of one, or more bubbles within the liquid (Oguz & Prosperetti 1990). At very high-velocity droplet impact, the rising cylindrical sheet at the upper edge of the crown bends inward as surface tension pulls on it, forming a canopy. This canopy then collapses towards the centre, generating a downward spiral jet that propagates towards the bottom of the cavity (Wang *et al*. 2023).

Indeed, the hydrodynamics of the cavity depend on the impact parameters, i.e., impact velocity, force of gravity, droplet size, and physical properties such as viscosity, density, and surface tension. Depending on these parameters, various regimes viz. floating, bouncing, coalescence, and splashing may arise (Rein 1996). Our present research is focused on the splashing regime. Research has been devoted to the splashing regime for Newtonian liquids, especially towards the scaling of the maximum cavity size, the time to reach maximum cavity depth (Engel 1966; Leng 2001), the spatio-temporal evolution of the cavity (Engel 1967;

Morton *et al*. 2000), the growth and disintegration of the crown (Okawa *et al*. 2006; Zhang *et al.* 2010), the generation and breakup of the central jet (Fedorchenko & Wang 2004; Ray *et al.* 2015), and underwater acoustic signatures of the impact (Prosperetti & Oguz 1993). Also, the effects of droplet oblique-impact (Gielen *et al.* 2017), liquid immiscibility (Lhuissier *et al.* 2013; Jain *et al.* 2019) and visco-plastic effects (Jalaal *et al.* 2019) on impact hydrodynamics have been studied.

After an exhaustive review of literature, it is evident that despite extensive research on Newtonian droplet-pool interactions through experimental, numerical, and analytical studies, explorations and understanding the effect of rheology on the post-impact hydrodynamics of droplet-pool interactions remains largely scarce, especially in the context of elastic liquids. The presence of trace amounts of very long-chain polymers in a Newtonian solvent renders the resulting solution elastic (Boger fluids), which significantly influences the cavity formation and collapse, crown dynamics, and jet ejection and disintegration processes, owing to the drastically altered shear and extensional rheology. We often encounter viscoelastic/elastic fluids such as human mucus, paints, inks, gums, resins, medicinal gels, microbial films, and industrial thickeners. The temporal evolution of viscoelastic/elastic liquid filaments, and formation of daughter droplets has significance in systems ranging from additive 3D printing, to arrest of droplet rebound on superhydrophobic surfaces (Dhar *et al.* 2019; Kamaluddin *et al.* 2024 & 2025), to the controlled deposition of cells in tissue engineering (Basaran *et al.* 2013). Understanding the fundamental mechanisms of the fragmentation process influenced by viscoelastic/elastic effects may facilitate the control of splashing and instabilities associated with droplet impact processes.

Experimental, theoretical, and computational studies have been reported in literature towards understanding the physical behavior underlying the thinning, and breakup of slender viscoelastic-liquid filaments (Clasen *et al.* 2006; Eggers & Villermaux 2008; Ardekani *et al.* 2010; Turkoz *et al.* 2018; Bera *et al.* 2026). A typical theoretical framework for studying viscoelastic filament thinning employs the slender jet profile formulation derived using the lubrication approximation, coupled with constitutive equation of the viscoelastic model (Eggers & Villermaux 2008; Clasen *et al.* 2006). This approximation reduces the problem to a one-dimensional description of the jet radius, axial velocity, and polymeric stresses. The self-similar profiles of a viscoelastic filament undergoing thinning have been computed using direct numerical simulations (Turkoz *et al.* 2018; Snoeijer *et al.* 2020), revealing the significant local influence of additional polymeric-stress components near the final pinch-off.

During the breakup process, as the system approaches the Newtonian pinch-off time, the polymer chains undergo rapid stretching, leading to a substantial increase in elastic stress, which becomes sufficient to balance the diverging capillary pressure in the elasto-capillary regime. Consequently, the minimum neck radius decays exponentially with time, a behavior that has been observed both experimentally (Bazilevsky *et al.* 1990; Entov *et al.* 1997; Amarouchene *et al.* 2001; Clasen *et al.* 2006; Deblais *et al.* 2020) and from numerical simulations (Bousfield *et al.* 1986; Etienne *et al.* 2006; Bhat *et al.* 2010; Morrison & Harlen 2010; Turkoz *et al.* 2018; Eggers *et al.* 2020).

The (semi)-analytical works employ the Oldroyd-B (Bird *et al.* 1987; Clasen *et al.* 2006) constitutive model, accounting for the infinitely extensible polymer molecules dissolved in a viscous solvent, while the final breakup of fluid thread has been investigated through numerical simulations employing the FENE-P model, considering finite extensibility of polymer chains to evaluate the polymeric stress evolution (Anna *et al.* 2001; Fontelos & Li 2004; Wagner *et al.* 2005). The thinning, breakup, and subsequent formation of satellite droplets in polymeric liquid filaments have been investigated in canonical configurations, such as liquid jets issuing from circular nozzles (Eggers 1997; Zinelis *et al.* 2024; Clasen *et al.* 2009; Ardekani *et al.* 2010; Clasen *et al.* 2006, Deblais *et al.* 2020; Goldin *et al.* 1969), and liquid bridges stretched between two end-plates. In such systems, the flow conditions are externally controllable through either the imposed nozzle flow rate, or the stretching force applied to the endplates, enabling a relatively simplified description of the necking dynamics. In these configurations, the position of the minimum neck diameter remains independent of the axial coordinate, and the filament evolution is commonly analysed solely as a temporal (0+1) problem within a Lagrangian framework.

In contrast, the dynamics of cavity formation, its growth, stall, and collapse, the consequent Worthington jet-stretching, filament formation, and eventual breakup after impact of polymeric droplet on a deep polymeric pool remains very scarcely explored. Unlike conventional filament-thinning geometries, such a problem is intrinsically unsteady and spatially evolving, arising from the collapse of capillary waves at the base of the formed cavity. The emerging jet dynamics are governed by the impact velocity in conjunction with the propagation speed of the capillary waves along the cavity wall, where the latter is strongly influenced by polymer concentration and viscoelastic effects. Consequently, the minimum neck diameter is no longer confined to a fixed axial location; but instead, it evolves simultaneously in a coupled spatio-temporal (1+1) description of the necking and breakup dynamics. Recently, numerical simulations have been carried out to investigate bubble bursting in elasto-viscoplastic (Balasubramanian *et al.* 2024) and viscoplastic (Sanjay *et al.* 2021) media. These studies show how the interplay between elastic stress relaxation and yield stress profoundly alters the transient cavity collapse events and the subsequent formation and evolution of the jet. However, the physics of these events for elastic fluids (Boger fluids), which is the subject matter of our investigation, remains completely uncharted territory in literature.

To analyze the transient flow dynamics and the influence of viscoelastic behavior of polymeric liquids, the quantification of the extensional rheological properties of this class of fluids is necessary. In particular, the Capillary Breakup Extensional Rheometer (CaBER) (Bazilevsky *et al.* 1990; Yesilata *et al.* 2006) and Rayleigh Ohnesorge Jetting Extensional Rheometer (ROJER) (Keshavarz *et al.* 2015) are used to measure the extensional rheological properties of dilute or semi-dilute polymer liquids. In recent developments, instruments such as Dripping-onto-Substrate (DoS) rheometer is applied to measure the elasticity of biological fluids (Martínez Narváez *et al.* 2021; Zinelis *et al.* 2024).

As a closure to this section, the dynamics of cavity formation, its growth, stall, and collapse, the consequent Worthington jet-stretching, filament formation, and eventual breakup after impact of elastic (Boger) fluid droplets on deep elastic (Boger) fluid pool remains an unexplored topic of fundamental and intellectual curiosity. This forms the objective and motivation for our study on the elasto-hydrodynamics of droplet-pool interactions for elastic fluids (we use aqueous solution of very long chain polymers). Devoid of the viscous effects creeping into the physics at play, elastic fluids are expected to paint a clear illustration of the elasto-hydrodynamic effects in a pure fashion, which we believe holds tremendous significance towards understanding facets of non-Newtonian fluid dynamics. From this point forward, our paper is organized as follows: In § 2, we describe the experimental procedures and computer simulation methodology. In § 3, we show the morphologies of the cavity and jet evolution obtained from the different combinations of elastic fluid droplet-pool impact from our experiments, discuss the mechanisms observed, and summarize in a regime maps. In § 4, we put forward an energy conservation model for cavity radius evolution, to quantify elastic energy storage, and develop a theoretical model for the thinning dynamics of the elongated Worthington jets. In § 5, we validate our simulations using our experiments, present the evolution of the elastic stress tensor in the liquid filament, and the velocity field during cavity and jet evolution from the numerical analysis. The § 3-5 all aid in further describing and explaining the physics of the events observed and established in § 2. The paper ends with § 6, with the summary and conclusions drawn from our research.

## 2. Experiments and simulations

### 2.1. *Experimental materials and methods*

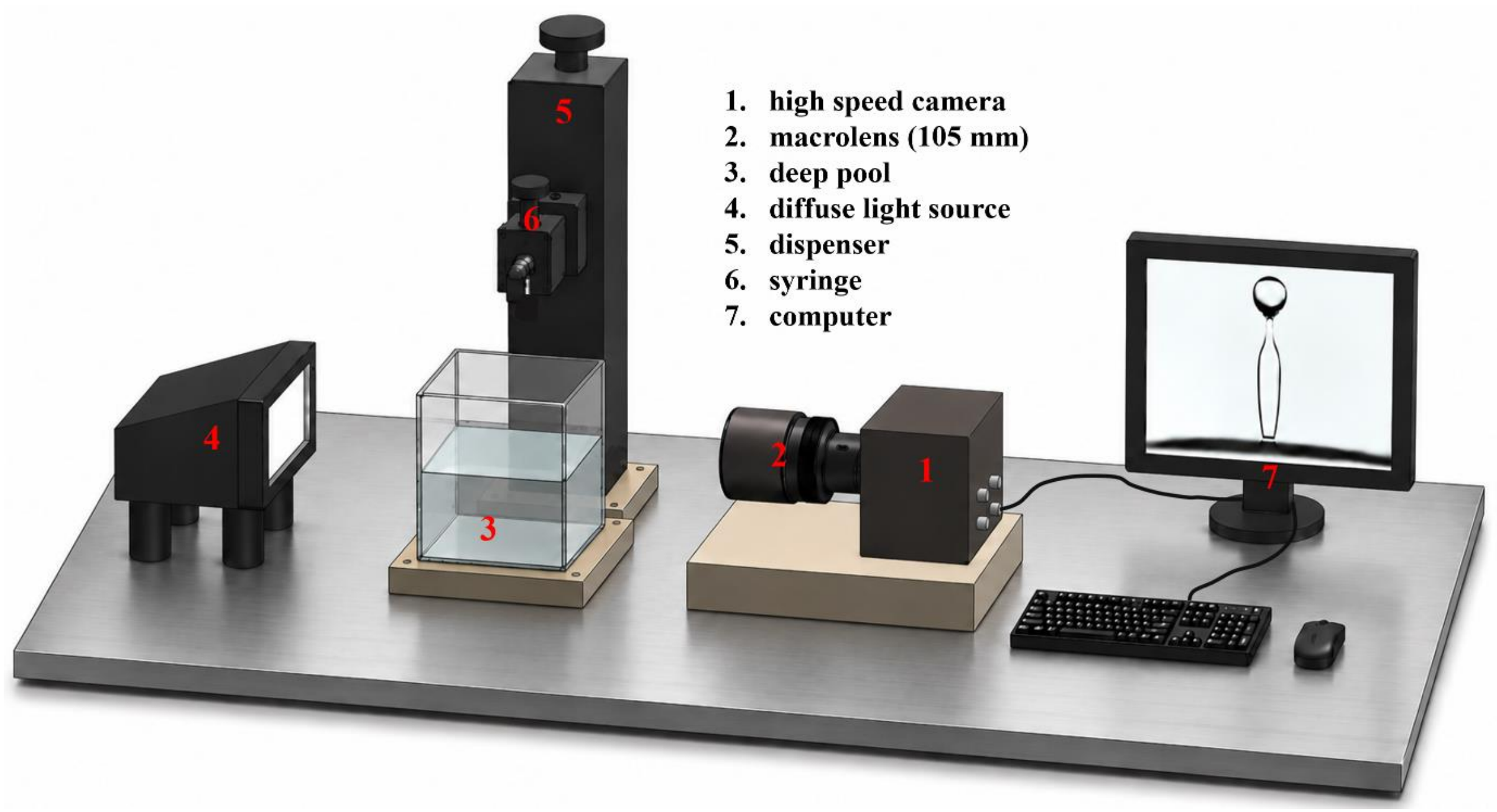

**Figure 1. Schematic of the experimental setup. The major components are labelled.**

Figure 1 illustrates the schematic of our experimental setup. A transparent container of pristine acrylic, of dimensions 70 mm × 70 mm × 100 mm was used to contain the deep pool. The container size was large enough to minimize all wall effects, as observed during trials. The droplets were quasi-steadily generated using a syringe and a digitized dispensing system. The metallic needle (flat head) used had a nominal diameter of 0.8 mm. We allow the droplet to grow slowly, until the weight of the droplet exceeds the surface tension force, causing it to detach from the needle tip with negligible initial velocity. A high-speed camera (Photron, UK) connected to a 100 mm macro lens (Nikon) captured the images at 4000 frames per second. A 100 W continuous white-light was used for illumination. The images were analyzed using Image-J software. We use aqueous solutions of very long-chain polyethylene oxide (PEO) (Sigma Aldrich) (Mol. Wt. ~5-6 million Da) at concentrations of 0, 250, 500, 1000, and 2000 parts-per-million (ppm) for the droplet and pool fluids. We use the extensional rheological properties reported by Sen *et al.* 2022, as the same PEO with a molecular weight of ~5-6 million Da and identical concentrations (ppm) is considered in our experiments.

The required PEO mass was measured using a precision electronic balance (accuracy of 0.1 mg) and each solution was stirred slowly with a magnetic stirrer for 24 hr to ensure complete dissolution and homogeneous concentration before use. Each experiment was repeated at least thrice to ensure reproducibility. The surface tension of the PEO solution for each concentration is $\sigma_P = 0.06 \pm 0.01$ N m$^{-1}$. Previous studies (Tirtaatmadja *et al.* 2006) have shown that addition of PEO reduces the surface tension of pure water from ~0.072 to ~0.06 N m$^{-1}$. However, once trace amount of PEO is added, further increase in PEO concentration does not change the surface tension of the resulting solutions. The polymer relaxation time ($\lambda$) and the shear viscosities of the polymer ($\mu_P$) and solvent ($\mu_s$) are taken from Sen *et al.* 2022, and are listed in table I. As seen from table I, while the viscosity of the fluids remains very similar to water (especially up to 1000 ppm), the relaxation times increase significantly, thereby establishing that the fluids are elastic Boger fluids. The height of the needle tip from the free surface of the pool was varied to obtain different impact velocities. Assuming free fall and neglecting aerodynamic drag, the impact velocity was calculated as $V_o = \sqrt{2gh}$ where, *h* is the vertical distance between the needle tip and the undisturbed pool surface. In our experiments, the impact velocities were varied in the range 1.5–4.3 m s$^{-1}$.

**Table I. Rheological and physical properties of the aqueous PEO solutions and control fluids used in the present study.**

| Polymer concentration, $C$ [ppm] | Viscosity, $\mu$ [mPa.s] | Surface tension, $\sigma$ [ N m$^{-1}$] | Density, $\rho$ [kg m$^{-3}$] | Relaxation time, $\lambda$ [ms] | $\mu_P / \mu_s$ |
|---|---|---|---|---|---|
| 0 | 1.00 | 0.071 | 1000 | ~0 | 0 |
| 250 | 1.01 | 0.060 | 1000 | 0.069 | 0.006 |
| 500 | 1.05 | 0.060 | 1000 | 0.137 | 0.013 |
| 1000 | 1.15 | 0.060 | 1000 | 0.192 | 0.281 |
| 2000 | 1.38 | 0.060 | 1000 | 0.362 | 0.533 |
| water-glycerol | 1.40 | 0.070 | 1038 | ~0 | 0 |

The impact dynamics of a polymer liquid droplet onto a deep polymeric-liquid pool depends on its radius $R_o$ and impact velocity $V_o$, solvent viscosity $\mu_s$ and zero shear rate viscosity $\mu_p$, surface tension $\sigma$, relaxation time $\lambda$, and gravitational acceleration $g$. These parameters yield the following set of five non-dimensional numbers which dictate the dynamics of the events:

$$We = \frac{\rho V_o^2 R_o}{\sigma},\ Bo = \frac{\rho g R_o^2}{\sigma},\ Fr = \frac{V_o^2}{g R_o},\ Oh = \frac{\mu_s}{\sqrt{\rho \sigma R_o}},\ De = \frac{\lambda}{t_c} \tag{2.1}$$

The Deborah number $De$ is the ratio of the polymer relaxation time to the inertia-capillary timescale $t_c = \sqrt{\frac{\rho R_o^3}{\sigma}}$. The Ohnesorge number $Oh$ compares the visco-capillary timescale and the inertio-capillary timescale. Both $Oh$ and $De$ play important role in determining the jetting and pinch-off processes. Since $Oh < 0.0046$ in all our experiments, the effect of dynamic viscosity on the breakup process is negligible. Herein, the Froude number $Fr$ represents the ratio of the kinetic energy of the impacting droplet to its gravitational potential energy immediately prior to impact. The Weber number $We$ is the ratio of the inertia to surface tension forces. The Bond number $Bo$ is the ratio of gravity forces to interfacial tension at the liquid air-interface. We fix the Bond number in our study at $Bo$ =0.27 (<1) to ensure that the pre-impact droplet remains spherical.

## 2.2 *Computational methodology*

Figure 2 shows the computational domain, and boundary conditions used in our simulations. The domain is 30 mm×50 mm in size, and a droplet of radius 1.5 mm is initially positioned 0.3 mm above the free surface of the pool. The droplet size is identical to that used in the experiments. The z-axis represents the axis of symmetry, along which gravity acts. The bottom and right boundaries are treated as walls with no-slip condition (mimicking the container inner-walls), while the top boundary is specified as an open boundary, allowing air to exit or enter the domain. The droplet is initialized with an impact velocity of 4.3 m $s^{-1}$, matching the experimental conditions corresponding to *We*=384. In our study, we couple the FENE-P constitutive model with the Cahn-Hilliard phase-field model for analysing the elastic fluid droplet-pool impact interactions. In the FENE-P model, a polymer molecule is assumed as two beads connected together by a finitely extensible Hookean spring (Bird *et al.* 1987).

Therefore, it is often applied accurately for the modelling of dilute polymer solutions, in which interaction between polymer molecules is negligible. As a thorough survey of literature suggests, this may well be the first attempt to simulate elastic fluid droplet-pool impact using the phase-field formalism.

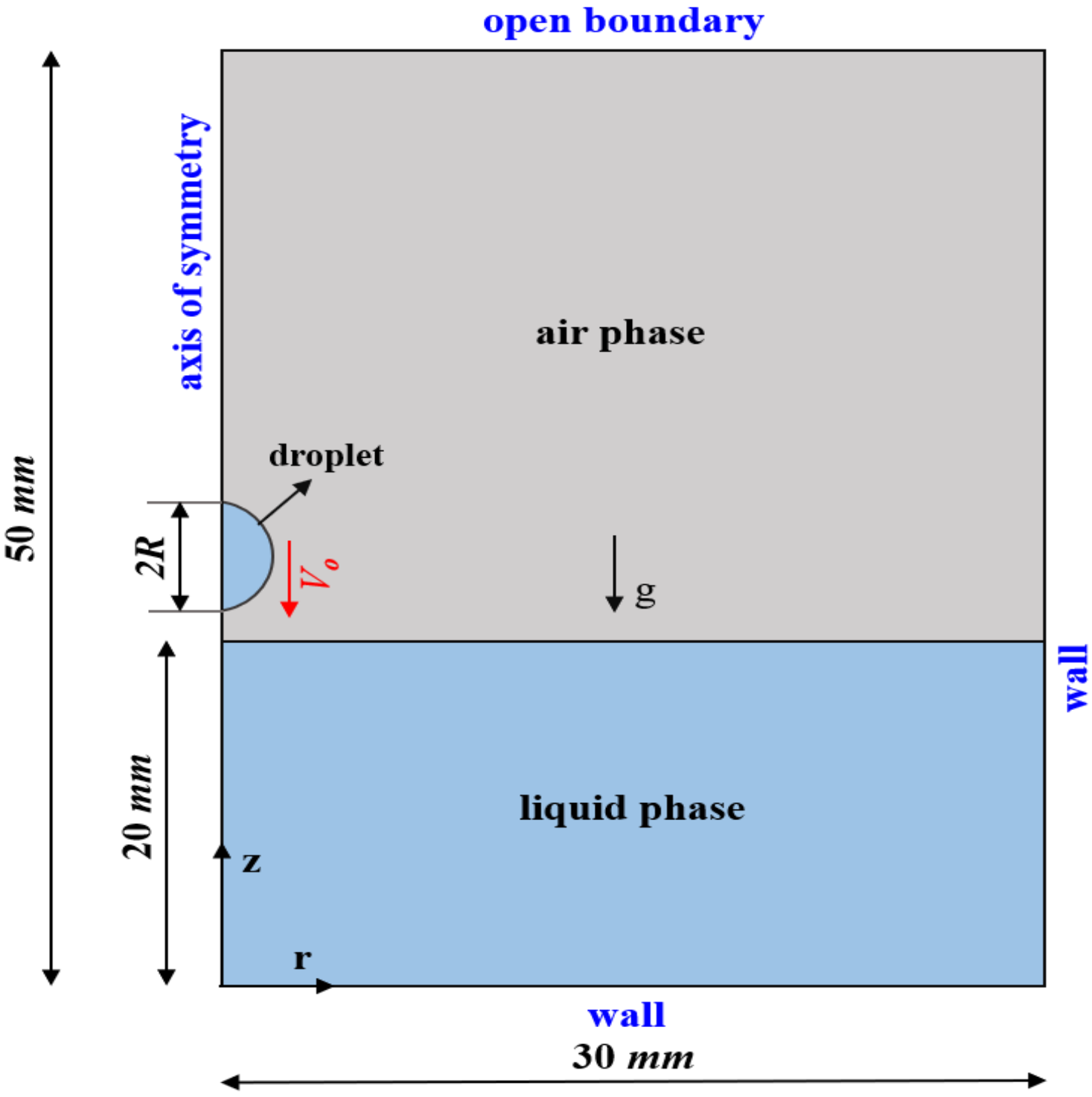

**Figure 2. Computational domain and the boundary conditions. All essential components are labelled.**

The convective Cahn-Hilliard (CH) equation (Cahn & Hilliard 1958; Cahn & Hilliard 1959) which describes the evolution of the order parameter $\phi$, minimizing the free-energy at the interface of two bulk phases, is expressed as

$$\frac{\partial \phi}{\partial t} + \nabla \cdot (\phi \mathbf{V}) = M \cdot \nabla^2(\mu) \tag{2.2}$$

Here, $\mathbf{V}$ denotes the velocity field vector, and $M > 0$ is known as the mobility or Onsager coefficient, which regulates the diffusion across the interface. The chemical potential, $\mu$, is determined by taking the derivative of the free energy with respect to $\phi$. The CH equation captures the temporal variation in $\phi$, accounting for the effects of advection from the velocity field. In our simulations, $\phi = 1$ represents the liquid phase, and $\phi = -1$ indicates the air phase. The CH equation is next intricately coupled to the FENE-P constitutive equation, via the introduction of continuum interface force term $\mu\nabla\phi$. The governing equations solved are as follows:

$$\rho\left(\frac{\partial \mathbf{V}}{\partial t}+\mathbf{V}\cdot(\nabla\cdot\mathbf{V})\right)=-\nabla p+\nabla\cdot\boldsymbol{\tau}+\rho\mathbf{g}+\mu\nabla\phi \tag{2.3}$$

$$\nabla\cdot\mathbf{V}=0 \tag{2.4}$$

Where, $\boldsymbol{\tau}$ represents the total extra stress tensor and $\mathbf{g}$ corresponds to the uniform gravitational acceleration.

The total extra stress is expressed as the sum of the viscous stress contribution from the Newtonian solvent liquid, $\boldsymbol{\tau}_s$, and the polymeric stress contribution arising from the presence of polymer molecules in the system, $\boldsymbol{\tau}_p$, as

$$\boldsymbol{\tau}=\boldsymbol{\tau}_p+\boldsymbol{\tau}_s \tag{2.5}$$

$$\boldsymbol{\tau}_s=\mu_s\left(\nabla\cdot\mathbf{V}+\nabla\cdot\mathbf{V}^{\mathrm{T}}\right) \tag{2.6}$$

The upper convective FENE-P constitutive equation in stress tensor form is expressed as (Bird *et al.* 1987)

$$\boldsymbol{\tau}_p+\left(\frac{\lambda}{Z(\tau_p)}\right)\overset{\nabla}{\boldsymbol{\tau}}_p=a\mu_p(\nabla\cdot\mathbf{V}+\nabla\cdot\mathbf{V}^{\mathrm{T}})\left(\frac{1}{Z(\tau_p)}\right)-a\mu_p\frac{D}{Dt}\left(\frac{1}{Z(\tau_p)}\right)\mathbf{I} \tag{2.7}$$

Where, $\tau_p\equiv\mathrm{tr}(\boldsymbol{\tau}_p)$, $Z(\tau_p)\equiv a+\frac{\lambda}{b\mu_p}\mathrm{tr}(\boldsymbol{\tau}_p)$ and $a\equiv\frac{b}{b-3}$. Here, $\mu_s$ denotes the solvent viscosity, whereas $\mu_p$ represents the zero-shear rate viscosity, and $b$ and $\lambda$ denote the finite extensibility parameter and relaxation time, respectively.

The simulations were performed using the commercial finite element method-based solver, COMSOL Multiphysics. The spatial derivatives in the governing equations were discretized using the finite element method, transforming the system into a set of ordinary differential equations, which were subsequently integrated over time to capture the evolution of the phases and the flow fields. Temporal integration was carried out using the implicit Backward Differentiation Formula (BDF), in conjunction with the MUMPS direct linear solver. The BDF employs variable-order backward differentiation schemes, typically ranging from first to fifth order accuracy, enabling efficient and stable time-marching. In our simulations, a second order BDF time-stepping scheme is employed, with the maximum Courant number limited to ~0.3. Linear finite elements were adopted for both the velocity components and the pressure field owing to their computational efficiency, and their ability to suppress spurious numerical oscillations, thereby improving the overall stability of the simulations. The computational domain was discretized using a non-uniform mesh, composed of triangular elements. The mesh was most-densely refined near the pool's free surface and the droplet-air interface, where the element size was at least ~100 times smaller than the droplet radius. A relative tolerance of 0.001 was applied to all dependent variables to ensure smooth and stable convergence of the numerical solution. Furthermore, isotropic diffusion stabilization with tuning parameter of 0.25 was employed to suppress any numerical inconsistencies. For the last phase of grid independence study, the number of mesh elements was varied from ~366196 to ~774722. The resulting changes in the radial and axial velocity components were noted to be less than 0.5%.

## 3. Experimental results and discussions

### 3.1 *Predominantly elastic effects driven hydrodynamics*

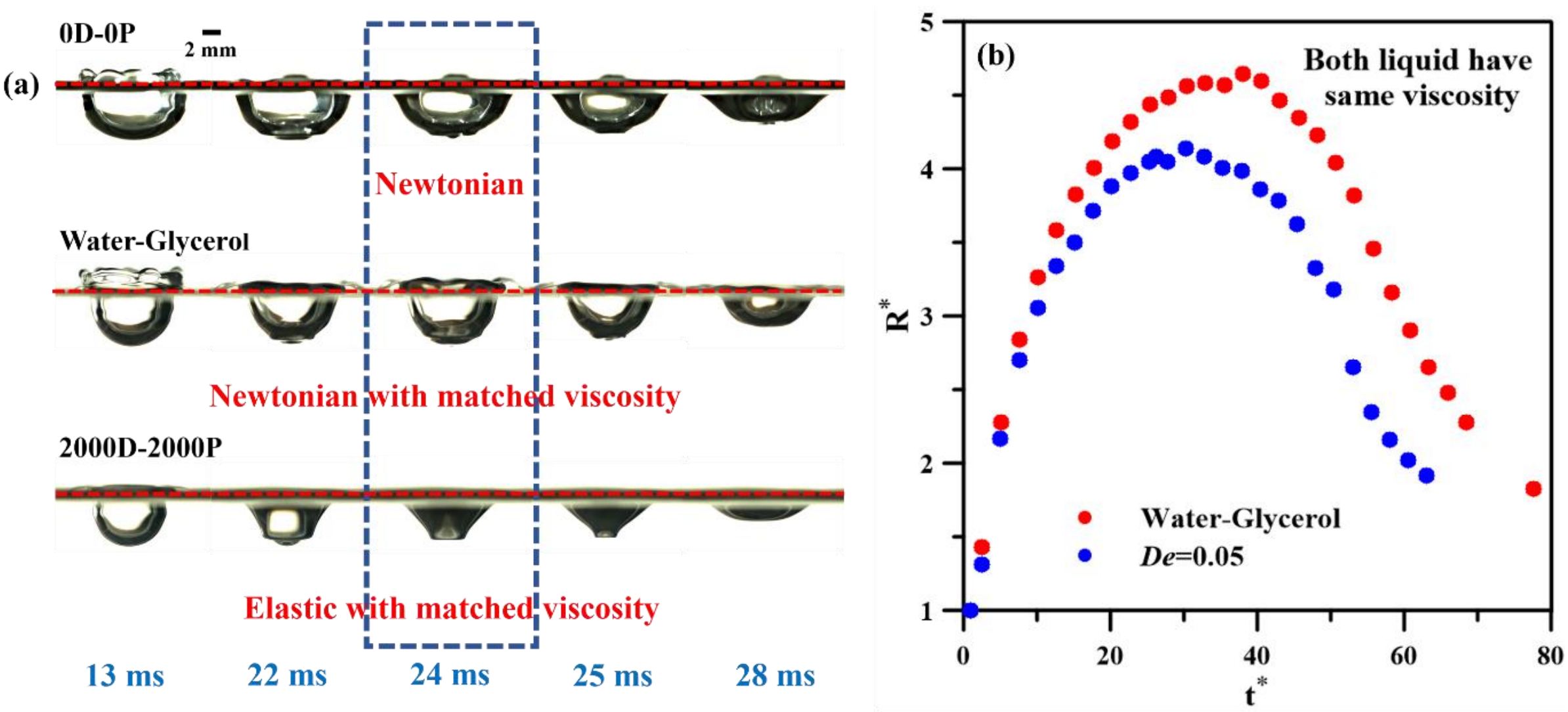

**Figure 3. Evidence of negligible intervention by viscosity (up to *De* =0.05) on the fluid dynamic events during elastic fluid droplet-pool interactions, (a) qualitative comparison of the cavity shapes, (b) temporal evolution of the cavity radius during the expansion and retraction phases for *De* =0 and *De* =0.05, at *We* =192. The differences observed are solely governed by fluid elasticity.**

Before the start of our discussions, the first thesis that needs to be established is whether the phenomena observed are driven by solely viscous effects, viscoelastic effects, or solely elastic effects, given that the fluids are non-Newtonian by nature. Figure 3a illustrates a comparative analysis of the cavity morphology resulting from droplet impact onto three distinct liquid pools; *top row*: a Newtonian fluid droplet (water) falling onto its own pool; *middle-row*: a Newtonian liquid droplet but with higher viscosity (water-glycerol solution) falling onto its own pool; and *bottom-row*: an elastic liquid droplet (2000 ppm aqueous PEO solution) falling onto its own pool. The water-glycerol solution (*middle row*) was prepared to possess the same viscosity as the 2000 ppm elastic fluid (*bottom row*). Hence, comparison of top and middle rows indicated the role of fluid viscosity, while comparing the middle and bottom rows indicate the role of fluid elasticity.

During the cavity evolution phase, the cavity shape for the water-glycerol solution (*middle row, 2$^{nd}$ to 4$^{th}$column*) remains hemispherical, mirroring the cavity shape observed in the case of pure water (*top row, 2$^{nd}$ to 4$^{th}$ column*). Here, only the cavity size in case of water-glycerol is smaller to some extent compared to the water case, as the viscosity of the water-glycerol is ~38% higher than water (see table 1), and hence the minor damping of cavity growth. In sharp contrast, the elastic fluid exhibits a distinctively trapezoidal cavity shape (*bottom row, 2$^{nd}$ to 4$^{th}$ column*), despite having the same viscosity as the water-glycerol solution. Also, the size of the 2000 ppm elastic fluid cavity is smaller to some extent than the water-glycerol case, despite the viscosities being same. This marked difference in cavity morphology and size; in two fluids whose shear viscosities are matched, confirms that the morphological alterations observed in our studies are a direct consequence of the polymer relaxation mechanisms, and predominantly the influence of elastic stresses. Hence our observations are dictated predominantly by elastic effects, and only very weakly by viscous effects (again this is only true for the 2000 ppm solutions; for the solutions up to 1000 ppm, the viscosity is higher than water at best within ~ 15%, and thus in those cases the effects observed would be purely liquid-elasticity driven).

Figure 3b illustrates the temporal evolution of the mean cavity radius generated by droplet impact onto two liquid pools of identical viscosity: a Newtonian water-glycerol solution droplet in its own pool, and an elastic 2000 ppm solution droplet in its own pool (corresponding to the middle and bottom rows of figure 3a). The initial cavity expansion (up to $t^*$ ~15) is nearly identical in both cases, confirming the dominance of inertia force during the early stages of cavity formation. But as the cavity evolves further, prominent differences in the dynamics appear during both expansion and retraction phases. Since the viscosities of the two liquids are same, and their surface tensions are very similar (see table 1), and all experimental conditions are identical, the observed changes in dynamics can be attributed exclusively to elastic stress distribution and relaxation processes within the elastic liquid.

From the velocity contours (see figure 19), it is interpreted that polymer chain stretching along the cavity walls is non-uniform. The polymer chains undergo greater degree of stretching near the cavity base, than along the surrounding cavity wall, leading to a non-uniform distribution of elastic stresses. This uneven stress distribution modifies, and delays the propagation of capillary waves along the cavity interface. During the onset of cavity retraction phase, pressure builds up near the cavity base, tending to drive the cavity floor upwards. Simultaneously, the concentrated elastic stresses near the floor oppose this motion. The resulting competition between the upward pressure force, and the resisting elastic stresses leads to flattening of the cavity floor. Meanwhile, the capillary waves continue to propagate towards the cavity's centre. However, their smooth convergence is hindered by the localized interplay between pressure and elastic stresses near the cavity floor, leading to the formation of a notch-shaped profile. Consequently, the cavity develops a flattened bottom, and inclined sidewalls, yielding trapezoidal cavity morphology than a hemispherical one in the absence of elastic effects. The detailed discussions on the change in cavity shape will be in subsequent sections.

### 3.2 *Effects of We and De when water droplet impacts upon elastic pools*

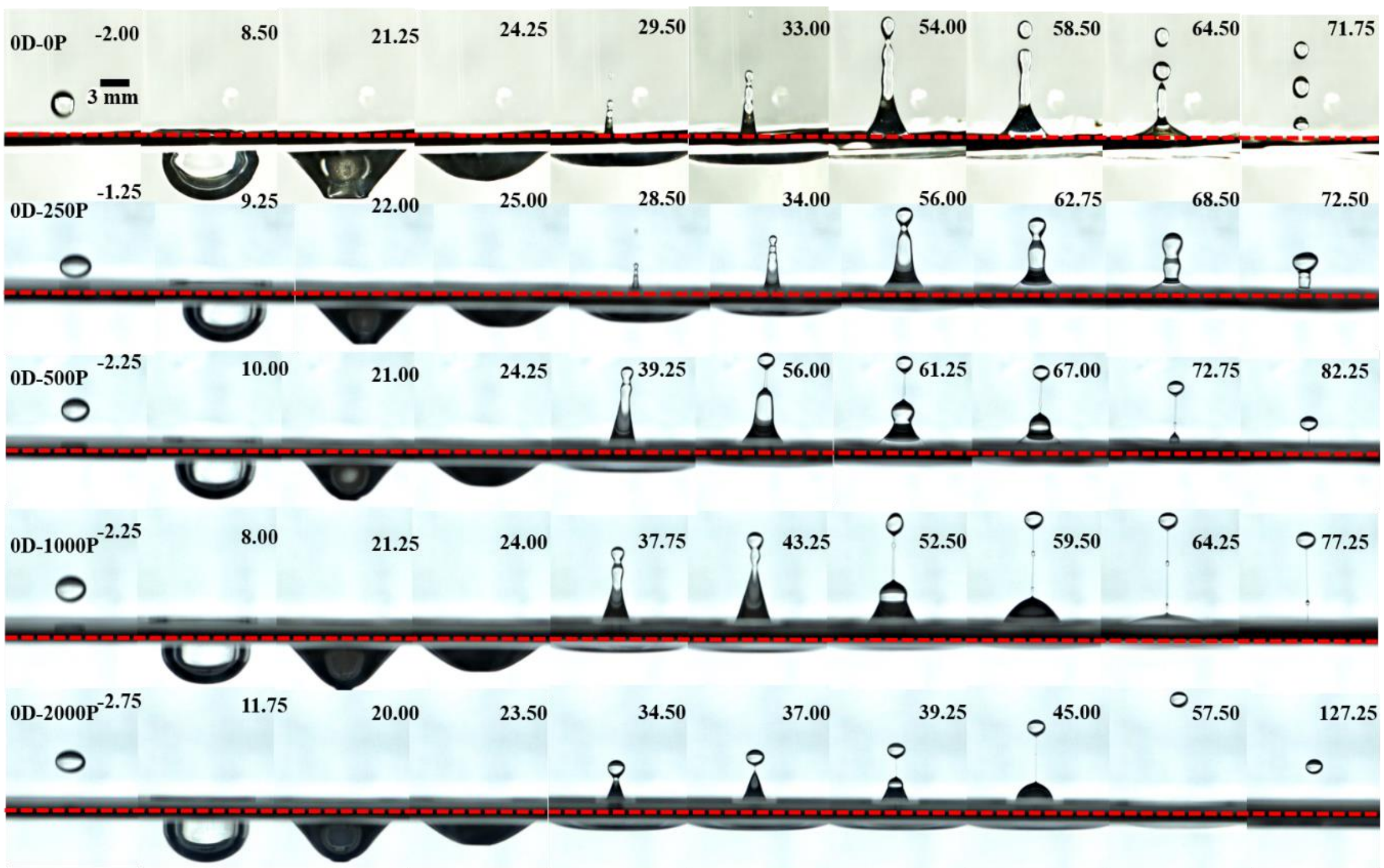

**Figure 4. Temporal evolution of cavity, and jet dynamics when Newtonian droplet (water, $De_D = 0$) impacts upon pools of different fluid elasticity (PEO solutions, from top to bottom, $De_P =$ 0, 0.01, 0.018, 0.026 and 0.05) at *We*=96. All times are in ms. The scale bar in row 1 column 1 is same for all the snapshots.**

Figure 4 illustrates the morphological transitions of the cavity, and of the jet, as the elasticity of the pool is increased (polymer concentration of pool varied from 0 to 2000 ppm), while droplet was maintained Newtonian (water). For brevity, hereafter the polymer concentrations of the droplet and pool are represented by their corresponding Deborah numbers, $De_D$ and $De_P$, respectively. From figure 4, it is evident that the cavity retains an approximately hemispherical shape during the expansion stage, for all cases considered. However, during the retraction phase, the cavity gradually transforms into a more trapezoidal shape. In the $De_P$ =0 case (figure 4, 1$^{st}$ row), capillary instability causes small axisymmetric perturbations on the cylindrical jet surface to grow. As the neck thins, the local jet curvature increases, leading to a rise in the capillary pressure. The resulting pressure gradient drives fluid away from the neck towards the neighbouring bulges, causing the neck to thin further. This process continues until the neck radius approaches zero, wherein the capillary pressure diverges and pinch-off occurs, resulting in the breakup of the jet into secondary droplets.

For $De_P$ =0.018 and 0.026 (figure 4, rows 3rd and 4th), secondary droplets remain connected by very thin filaments, forming the characteristic beads-on-a-string (BOAS) morphology of non-Newtonian fluids. As the filament thins, the polymer chains undergo substantial stretching, leading to a pronounced increase in the localized extensional viscosity. Consequently, elastic stresses developing within the filament become sufficiently large to locally counterbalance the rapidly increasing capillary pressure, which prevents filament rupture, and leads to very thin, but long-lived filaments.

A key observation here is that the BOAS develop only when the neck becomes asymmetric before the formation of the thin filament (figure 4, at 39.25 and 43.25 ms corresponding to the 3rd and 4th rows, respectively). In contrast, if the filament forms while the neck remains symmetric, bead formation does not occur, even if asymmetry develops at later stages. For $De_P$ =0.05 (figure 4, row 5th), no beads are observed despite the high polymer concentration, because the neck thins symmetrically and the filament is established before any significant asymmetry sets in (see the neck-thinning snapshots in the last row of figure 4).

If the growth rate of the disturbance (Rayleigh instability) of the liquid column exceeds the inverse of relaxation time ($\lambda^{-1}$), polymer chains begin to stretch during the early stages of the filament evolution, thereby suppressing bead formation (Wagner *et al.* 2005). A similar mechanism is observed in the present study. This can be seen by comparing the 5th, 6th and 7th columns of the 5th row in figure 4, with the corresponding events of the 3rd and 4th rows, illustrating the temporal evolution of the Worthington jet. For the 0D-2000P case (5th row), the filament forms at ~37 ms, whereas filament formation occurs quite later on, at ~ 44 ms, for the 0D-1000P (4th row) and 0D-500P (3rd row) cases. This indicates that the growth rate of the Rayleigh instability is much higher in the 0D-2000P case than in the other two cases. Hence, the growth rate of the disturbance exceeds the inverse of the polymer relaxation time, causing polymer stretching to commence during the initial stage of the filament evolution. As a result, the filament remains axisymmetric, preventing the development of the asymmetry required for the BOAS formation. In contrast, for the 0D-500P and 0D-1000P cases, the disturbance growth rate is comparable to, or smaller than, the inverse of the relaxation time. Consequently, polymer stretching is delayed until a later stage of the filament evolution, allowing asymmetry to develop, and ultimately leading to the formation of the BOAS. Furthermore, since the disturbance grows more rapidly in the 0D-2000P case, the polymer stretching takes place from initial motion. Therefore, the polymer molecules approach their finite extensibility limit, after which the polymer chains can no longer support extensional stresses, resulting in the final pinch-off and detachment of the filament from the droplet (see the final snapshot of the 5$^{th}$ row).

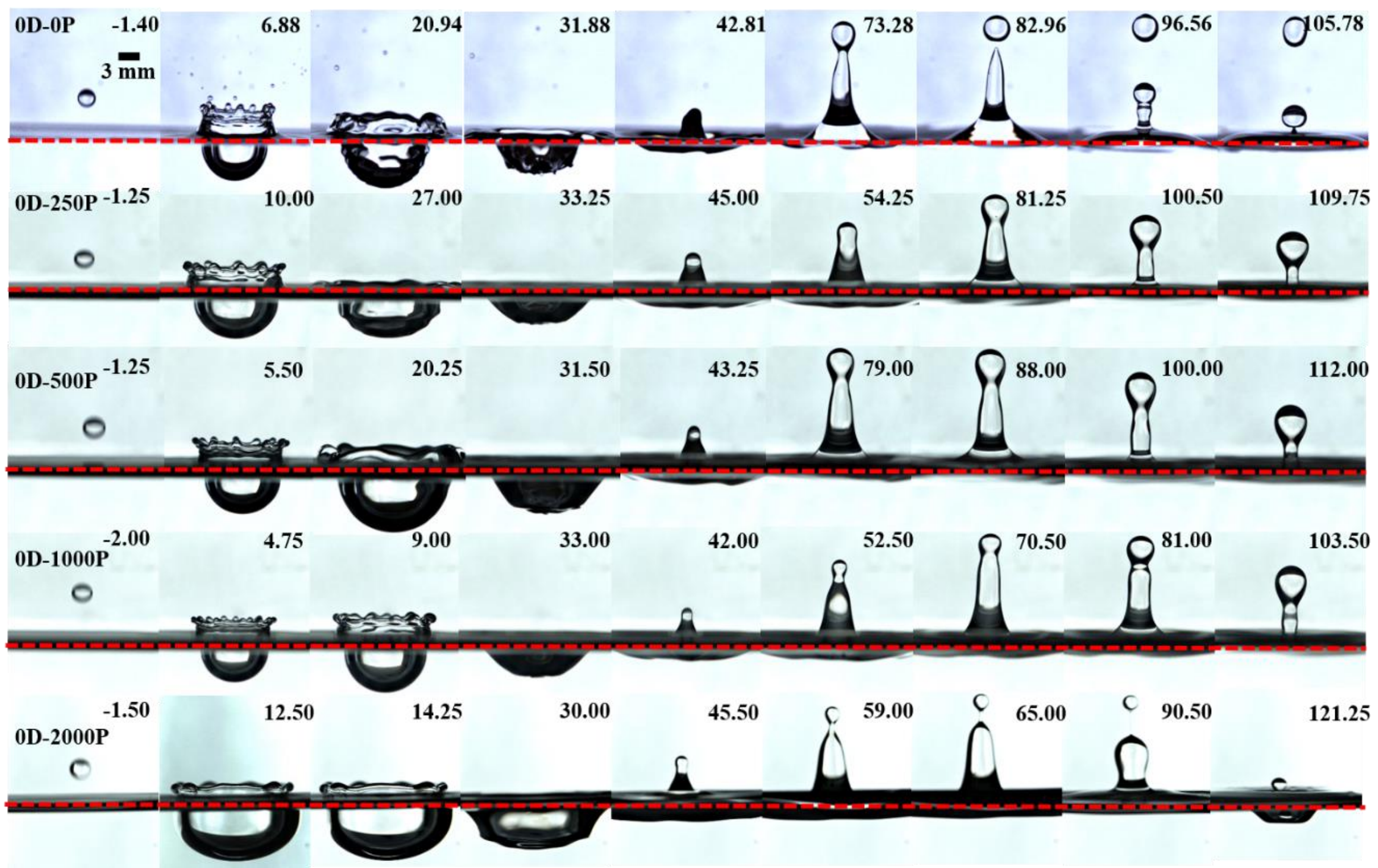

**Figure 5. Temporal evolution of cavity, and jet dynamics when Newtonian droplet (water, $De_D = 0$) impacts upon pools of different fluid elasticity (PEO solutions, from top to bottom, $De_P =$ 0, 0.01, 0.018, 0.026 and 0.05) at *We*=384. All times are in ms. The scale bar in row 1 column 1 is same for all the snapshots.**

Figure 5 illustrates the temporal evolution of the cavity and jet dynamics at a higher impact velocity (*We*=384) compared to figure 4. Due to high inertia, the cavity size grows to a larger extent, and the cavity penetrates further into the pool than at *We*=96 (figure 4). For $De_P$ =0, (figure 5, row 1), numerous tiny droplets are ejected from the destabilized cylindrical crown. The subsequent breakup of the Worthington jet, driven by the Rayleigh-Plateau instability, leads to the formation of secondary droplets. As $De_P$ increases, the crown instability is suppressed, and no pinch-off occurs at the jet up to $De_P$ =0.026. At $De_P$ =0.05, a satellite droplet remains connected to the primary jet by a very thin elastic filament. In contrast to *We*=96 (figure 4), the jet is thicker and more stable, and no BOAS is observed.

The overall jet breakup time increases with increasing $De_P$. The difference in jet breakup dynamics between the two impact conditions can be explained in terms of the elasto-capillary number ($Ec$), which represents the ratio of elastic stress to capillary pressure: $Ec = \frac{GR}{\sigma} = \frac{\mu_p R}{\lambda \sigma}$. In the $We$ = 384 case, the jet radius is larger than that for $We$ = 96, resulting in lower curvature ($1/R$) and hence smaller capillary pressure. Hence, the capillary pressure is insufficient to overcome the elastic stresses, and drive filament thinning toward pinch-off. Therefore, the jet stabilizes well before reaching the final pinch-off stage, preventing the formation of secondary droplets and a BOAS structure.

### 3.3 *Effects of We and De on impact dynamics when different elastic droplet- elastic pool combinations interact*

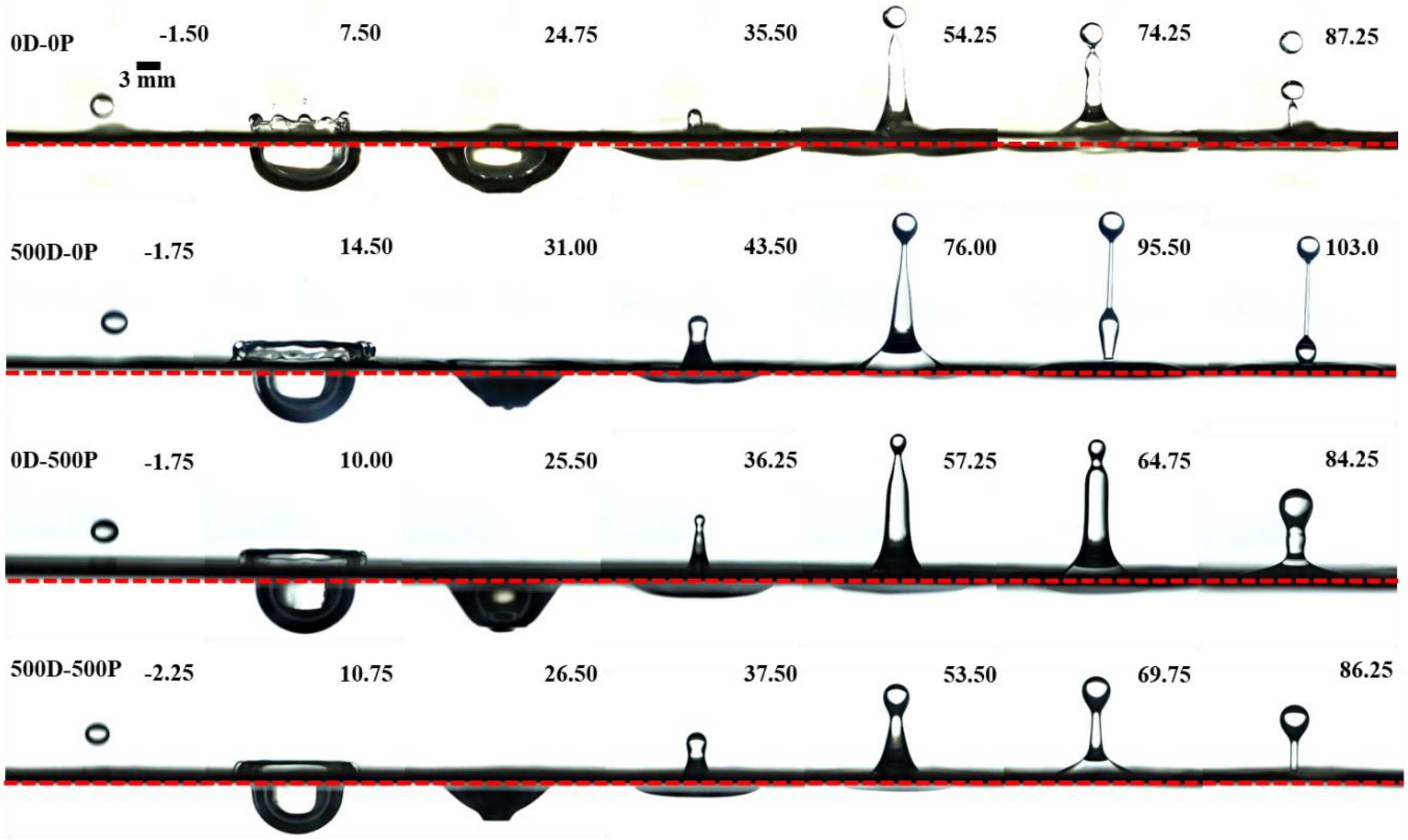

**Figure 6. Temporal evolution of cavity, and jet dynamics when different combinations of droplet ($De_D = 0$ and 0.018) and pool ($De_P = 0$ and 0.018) interact at *We*=192. All times are in ms. The scale bar in row 1 column 1 is same for all the snapshots.**

Figure 6 presents the temporal evolution of the cavity, and the jet at $We$=192, for different elasticity of the pool and droplet fluids. In the *1st row* (figure 6), the $De_P$ =0 droplet and $De_P$ =0 pool interaction produces a Worthington jet and secondary droplets. For $De_P$ =0 and $De_D$

=0.018 (*2nd row*), the jet emerges with a wide frustum-base, and tapers gradually towards the tip, producing a conical slender jet. The secondary droplet remains attached by a filament, while the jet continues to glide downwards along it. When $De_D$ and $De_P$ are both 0.28 (*4th row*), the cavity exhibits a trapezoidal reversal, the maximum jet height is lowest among all other cases shown here, and the secondary droplet stays connected to a comparatively thicker filament. Since the elastic stresses within the jet do not relax, they continuously counteract the capillary stresses, thereby preventing droplet detachment. As the elasticity of both the droplet and the pool is further increased (snapshots not shown for brevity), the maximum jet height decreases owing to the slower relaxation of elastic stresses, and the associated increase in elastic energy storage. As a result, jet development is delayed, a thicker filament is formed, and pinch-off is completely suppressed. During cavity formation, polymer stretching stores elastic energy that relaxes only partially during cavity collapse. The resulting non-uniform elastic stresses resist radial inflow, particularly near the cavity base, causing the collapsing cavity to transition from a hemispherical to a flatter, trapezoidal shape. With increasing $De_P$, capillary-wave propagation is delayed, reducing the maximum jet height. Further, elastic stresses oppose capillary-driven necking, thereby suppressing rapid pinch-off and producing thicker filaments.

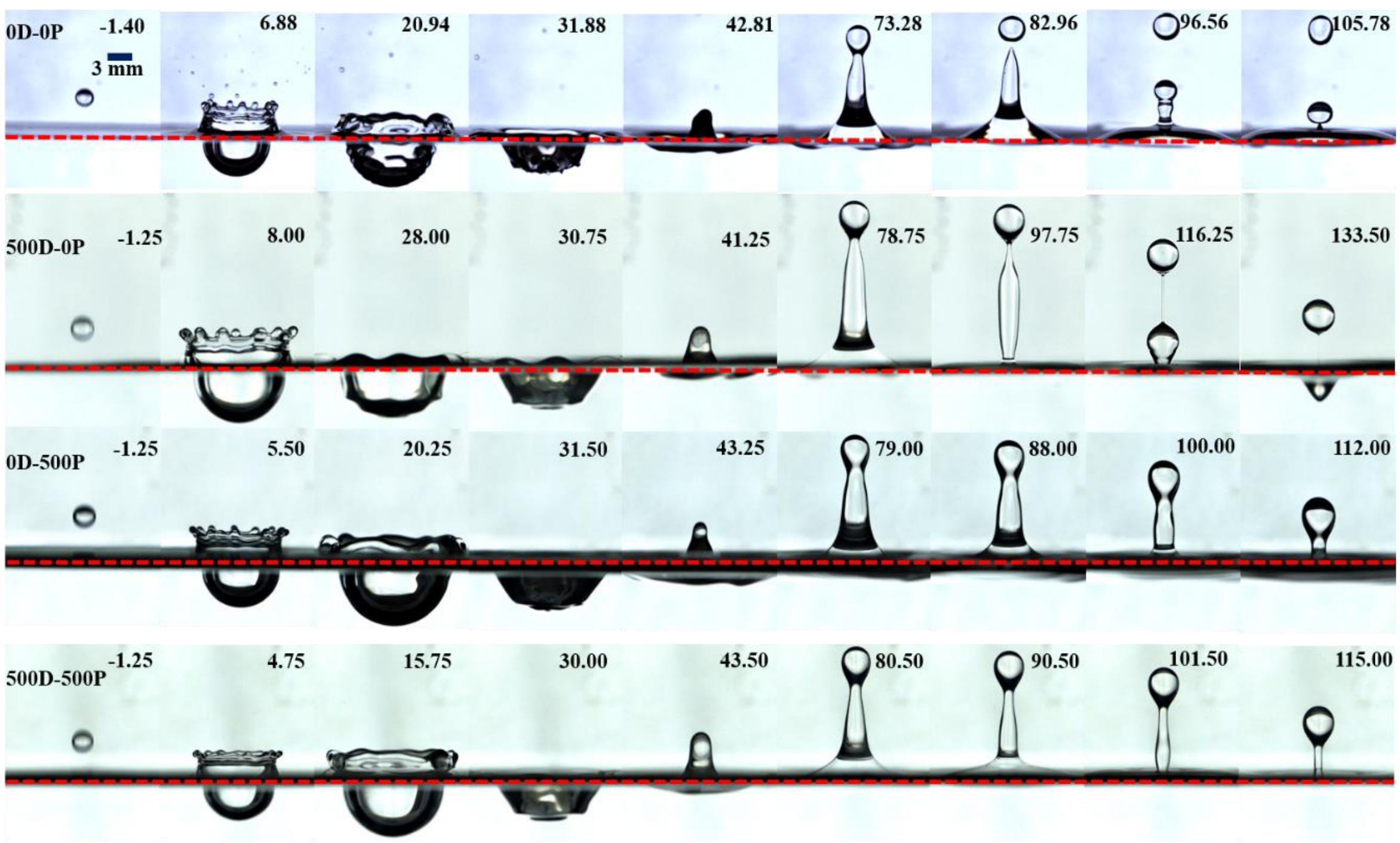

**Figure 7. Temporal evolution of cavity, and jet dynamics when different combinations of droplet ($De_D = 0$ and 0.018) and pool ($De_P = 0$ and 0.018) interact at *We*=384. All times are in ms. The scale bar in row 1 column 1 is same for all the snapshots.**

Figure 7 presents the temporal evolution of the cavity, and the jet at *We* = 384, for different combinations of pool and droplet elasticity. The formation of tiny droplets ejected from the periphery of the crown reduces, and eventually disappears as $De_P$ increases, owing to the greater elastic energy storage by the polymer molecules and suppression of capillary-driven instability. The cavity size decreases with increasing $De_P$. In the *1st row*, the $De_P$ =0 droplet and $De_P$ =0 pool interaction generates the Worthington jet accompanied by secondary droplets. As elasticity increases, the necking process is progressively suppressed by elastic stresses, leading to the formation of a stable filament. The secondary droplet remains attached to this filament without undergoing complete breakup. For the case For $De_P$ =0 and $De_D$ =0.018 (*2nd row*), the filament is thinner than that observed for $De_P$ =0.018 and $De_D$ =0.018 (*4th row*). The presence of polymer chains in both the droplet and the pool further delays pinch-off because elastic stresses more effectively compensate the capillary pressure. At this high *We*, the cavity shape during reversal remains similar across all cases due to the dominant inertial effect.

A comparison between the 0D-0P and 500D-0P cases at *We* = 192 and = 384 reveals a significant difference in the rate of neck evolution. At *We* = 192, pinch-off occurs at ~ 54 ms for the 0D-0P case, whereas thread formation occurs way later, at ~ 76 ms, for the 500D-0P case (figure 6, 5th event of 1st and 2nd rows). Similarly, at *We* = 384, pinch-off occurs at ~ 82 ms for the 0D-0P case, while thread formation is delayed until ~ 97 ms for the 500D-0P case (figure 7, 7th event of 1st and 2nd rows). These observations indicate that the growth rate of the Rayleigh instability is considerably lower in the 500D-0P case than in the 0D-0P case. Consequently, the growth rate of the disturbance is lower than the inverse of the polymer relaxation time, delaying the onset of polymer stretching until the later stages of neck deformation. As a result, the neck undergoes substantial capillary thinning before elastic stresses become significant, leading to the formation of a very thin filament. In contrast, for the other elastic cases, the growth rate of the Rayleigh instability is higher than in the 500D-0P case. Therefore, polymer stretching is initiated during the early stages of neck deformation, allowing elastic stresses to resist capillary thinning and resulting in the formation of a comparatively thicker filament.

**3.4** ***Regime maps for elastic droplet – elastic pool impact outcomes***

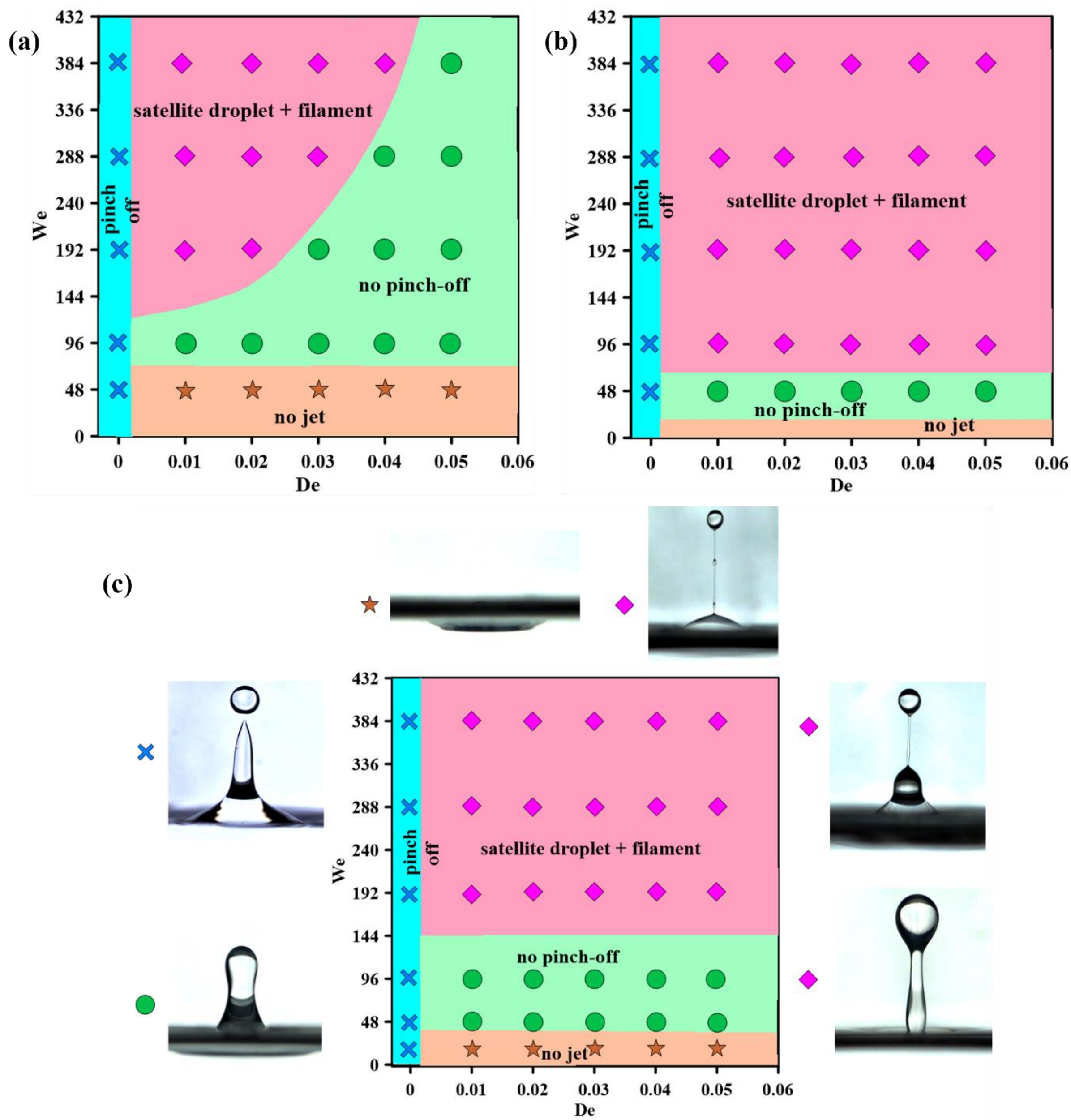

**Figure 8. Regime maps in the $[De, We]$ parameter space, highlighting four distinct regimes corresponding to '*pinch-off*', '*no jet*', '*satellite droplet+filament*', and '*no pinch-off*'. Here, $Bo$=0.27 is constant. Map corresponds (a) to droplet-pool systems with identical polymer concentrations, (b) to pure water droplets ($De_D$=0) impacting pools of different polymer concentrations, and (c) to droplets of varying polymer concentrations impacting a pure-water pool ($De_P$=0). The lines partitioning the regimes are the approximate crossover loci as obtained from our experiments.**

For figure 8, the experiments span over $De \in [0, 0.05]$ and $We \in [48, 384]$ while maintaining a constant $Bo = 0.27$, and $Oh \in [0.0031, 0.0046]$. The key dynamical features associated with different droplet-pool combinations yield three regime maps, as presented in figure 9. We identify four distinct regimes, viz. (i) breakup of Worthington jet tip into daughter droplet (*Pinch-off*), (ii) either the formation of a satellite droplet connected to the jet by an elastic fluid filament exhibiting BOAS, or a thin filament, or a thick filament (*Satellite droplet + filament*), (iii) no Worthington jet formation (*no jet*), and (iv) formation of a Worthington jet without subsequent disintegration (*no pinch-off*).

Figure 8a shows the regime map for droplets and pools of identical elasticity (polymer concentration). For a given *We*, the maximum cavity size decreases with increasing *De*, owing to the conversion of a portion of the impact kinetic energy into elastic energy stored by the stretching of polymer molecules concurrent with the cavity expansion. During cavity retraction, capillary waves converge at the cavity base, focusing the available kinetic energy to produce a Worthington jet. The subsequent evolution and breakup of the jet are governed by the interplay between elastic stress relaxation, capillary stress, and inertia, resulting in the different regimes of figure 8.

In the '*pinch-off'* regime (blue region, figure 8), the secondary droplet completely detaches from the jet. It is observed only for water droplets impacting a water pool, and occurs only for $De$=0 and $We \geq 48$. The breakup of the jet is driven by capillary forces. In the Newtonian system, no elastic stress is present to counteract the capillary pressure, which smoothly drive the breakup process, and pinch-off occurs even at low $We$. At very low $We$, no jet is formed for any droplet-pool combination, defining the '*no jet'* regime (brown region, figure 8). At low impact velocities or high elasticity, the propagation of capillary waves along the cavity wall is significantly slowed and delayed. As a result, either the kinetic energy, focused along the axis of symmetry by the converging capillary waves, is insufficient to initiate jet formation, or, the cavity returns to its equilibrium state before the capillary waves converge at the cavity base, thereby subduing any jet. This regime occupies the narrowest range (figure 8b) when $De_D$=0 and the pool elasticity is varied, whereas it extends over a broader range for identically elasticity droplet-pool combinations (figure 8a).

Upon increasing $De_P$ while maintaining a pure water droplet ($De_D$=0) (see figure 8b), a *no pinch-off* regime emerges for $De_P$=0.01-0.05 at $We$=48 . In this regime, a Worthington jet is formed but does not undergo breakup into a secondary droplet (green data points in figure 8b). This regime strongly depends on the polymer concentration in both the droplet and the pool. When both the droplet and the pool have a high polymer concentration ( $De = 0.05$ ), the jet remains intact even at high $We$=384. The generation of elastic stress during jet stretching continuously opposes the capillary stress responsible for necking and breakup. As the Deborah number of both the droplet and the pool increases, the elastic stress becomes sufficiently large that the capillary pressure is unable to overcome it and initiate jet breakup. Consequently, the jet elongates continuously without producing secondary droplets. A similar trend is observed when the droplet elasticity is increased while keeping the pool as pure water ( $De_P$ =0; figure 8c), where the *no pinch-off* regime expands over a wider range of $De_D$=0.01-0.05 and $48 \leq We \leq 144$.

The *satellite droplet+filament regime* is characterized by the formation of a satellite droplet that remains connected to the Worthington jet through a cylindrical liquid filament. Three distinct morphologies are identified within the *satellite droplet + filament* regime: (i) a satellite droplet connected by a polymeric thread exhibiting a beads-on-a-string (BOAS) structure, (ii) a satellite droplet connected by a thin filament, and (iii) a satellite droplet connected by a thick filament. In this regime, the breakup dynamics undergo a transition from the inertio-capillary regime to the elasto-capillary regime. During the early stage of jet evolution, elastic stresses are negligible, and the jet dynamics are governed primarily by inertia and capillary forces, which promote jet thinning and neck formation. As the filament continues to thin, the polymer molecules within the cylindrical filament undergo strong extensional stretching, leading to a rapid buildup of elastic stress. This increasing elastic stress counteracts the capillary pressure that drives necking. Eventually, a balance is established between the elastic stress and the capillary pressure (elastocapillary regime), stabilizing the satellite droplet and the connecting filament. For identical droplet-pool combinations (see the pink region in figure 8a); this regime is observed for intermediate Weber numbers and low Deborah numbers. Specifically, for $De \leq 0.04$ and sufficiently high $We$, jet breakup proceeds through the formation of a satellite droplet connected by an elongated filament. As $De$ is further increased at high Weber numbers, the system transits to the *no pinch-off* regime (green region in figure 8a), in which the Worthington jet no longer breaks up and filament formation is completely suppressed. For impacts of a pure water droplet onto an elastic pool, the *satellite droplet+filament* regime occupies the largest region of the regime map ( $De_P$ =0.01-0.05 and $We > 96$ ).

In the present experiments, the first signature of BOAS structures was observed at *We* =96 for $De_P$ =0.018 and $De_D$ =0 (see the snapshots at 72.75 ms corresponding to the 3rd row in figure 4) in the *satellite droplet+filament* regime (see the pink data points in figure 8c). The BOAS structure is a hallmark of the intermediate elasto-capillary thinning regime in extensional flows of viscoelastic fluids (Goldin *et al.* 1969; Clasen *et al.* 2006). In this regime, elastic stresses generated by stretched polymer molecules substantially delay capillary-driven breakup, thereby prolonging the lifetime of the slender filament. As the filament continues to thin, capillary pressure gradients transport fluid from the slender necks toward regions of lower curvature, giving rise to a series of spherical beads connected by thin viscoelastic filaments.

## 4. Theoretical formulation

### 4.1. *Cavity evolution dynamics*

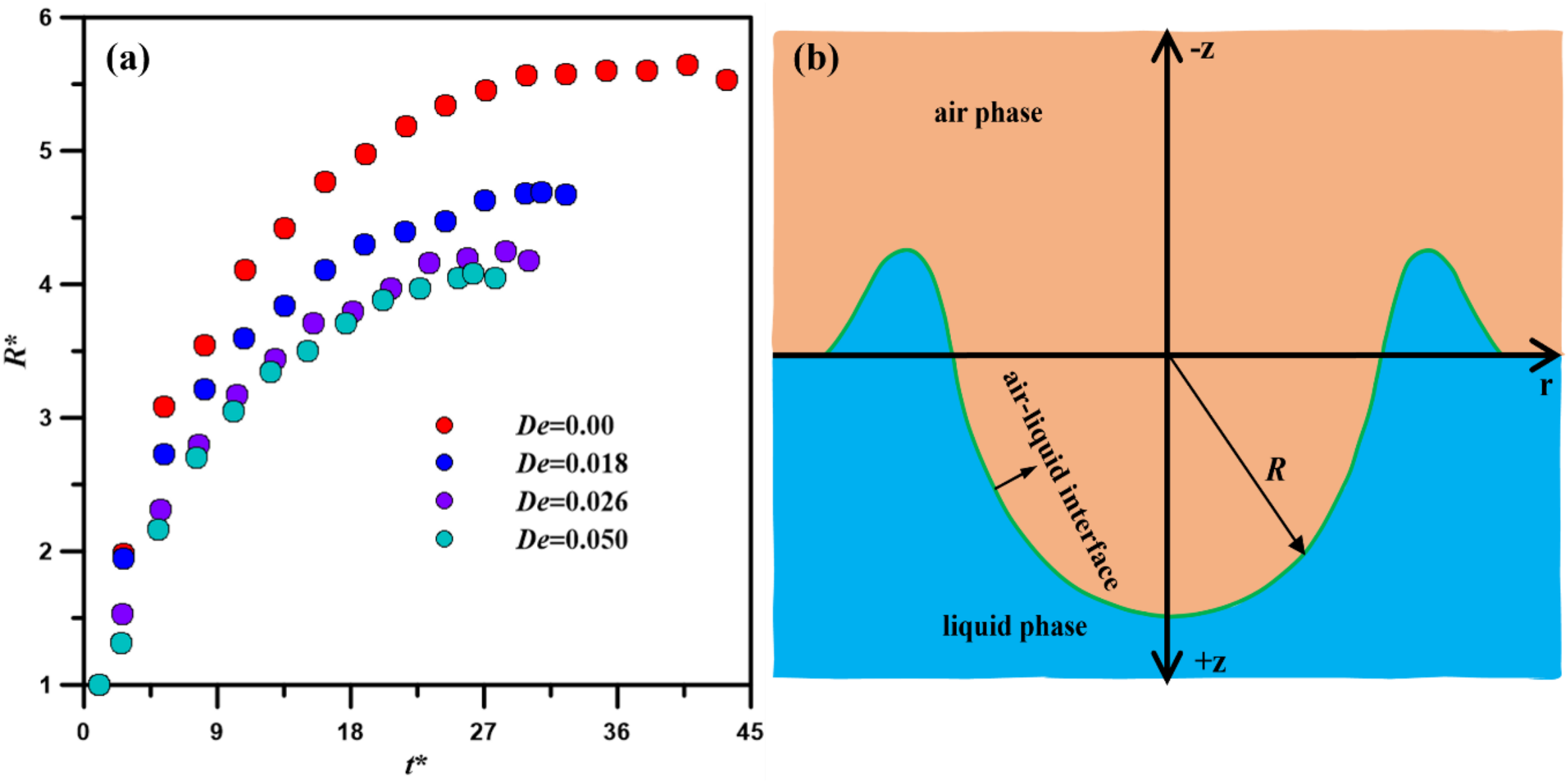

**Figure 9. (a) Evolution of the non-dimensional cavity radius (*R**) (from the experiments) with non-dimensional time (*t**) for different $De$, at $We$=192 ($t^* = 0$ indicates the instant of droplet-pool surface contact). The cavity radius is non-dimensionalised by the initial droplet radius. The evolution is shown only during the expansion stage. Here, the Droplet and the pool have same *De*. (b) Schematic of the cross-section of the expanding cavity in the $r$ - $z$ plane in aid of the theoretical model.**

Figure 9 presents the variation of non-dimensional cavity radius as a function of non-dimensional time. Here, $R^{*}=\frac{R}{R_o}$, Where $R_o$ is the initial droplet radius, and $t^{*}=\frac{tV_o}{R_o}$, where $V_o$ is the droplet impact velocity. Figure 9 shows that the cavity undergoes a nonlinear expansion behaviour while reaching its maximum radius. During the early stages of cavity growth ($t^{*}<10$), the data points for all $De$ values collapse; indicating that the cavity dynamics is primarily governed by inertial forces. For $De$=0, the cavity attains the largest expansion compared to all other cases. As the $De$ increases, a greater fraction of the droplet's kinetic energy is converted into elastic energy, stored within the polymer molecules which stretch along with the cavity due to shear (Bergeron *et al.* 2000; Samanta *et al.* 2013). Consequently, less energy remains available to drive cavity expansion, and suppresses its radial growth rate with increasing $De$. In the following sub-section, we quantify the amount of elastic energy stored by the polymer to better understand its influence on cavity dynamics.

**4.2 *Energy conservation model***

We quantify the amount of energy stored by stretching of the polymer chains during cavity expansion using energy conservation approach (Engel 1966, 1967; Lherm *et al.* 2022). The flow around the cavity wall is assumed to be incompressible and irrotational, and the cavity maintains an approximately hemispherical shape throughout its evolution. Since the cavity dynamics are predominantly inertia-driven, viscous dissipation is neglected. This assumption is further supported by figure 3, which indicates that viscous effects are negligible under the present conditions. Under these assumptions, the initial kinetic energy of the impacting droplet is redistributed into the gravitational potential energy, kinetic energy, and surface energy of the cavity, and a fraction is also stored as elastic energy due to the stretching of the long polymer molecules.

The cavity potential energy $U_P$ is the work required to lift the liquid contents of the cavity to the original undisturbed surface of the pool, against gravity. For a hemispherical cavity, $U_P$ is expressed as Eqn. 4.1(Engel 1966). Here, $R$ is the mean cavity radius, and $z$ is the depth.

$$U_P=\int \pi g\rho z^{2}rdr=\int_{0}^{R}\pi g\rho(R^{2}-r^{2})rdr=\frac{\pi g\rho R^{4}}{4} \qquad (4.1)$$

The surface energy of the cavity corresponds to the generation of new surface due to hemispherical cavity growth. It is equal to the difference of surface energy between the initial undisturbed planar surface area of the pool $\pi R^2$, and the cavity surface area $2\pi R^2$. The surface energy due to the formation of cavity is

$$U_S = (2\pi R^2 - \pi R^2)\sigma = \pi R^2 \sigma \tag{4.2}$$

The kinetic energy in the pool liquid around the cavity, below the original planar surface, is determined from the velocity potential. Due to symmetry, all variations in the azimuthal direction can be neglected. So, solving the Laplace equation in spherical co-ordinate, we obtain the radial velocity $u_r$ during cavity expansion, as

$$\nabla^2 \phi = 0 \tag{4.3}$$

$$\frac{1}{r^2}\frac{\partial}{\partial r}\left(r^2 \frac{\partial \phi}{\partial r}\right) = 0 \tag{4.4}$$

$$\phi = -\frac{c_1}{r} + c_2 \tag{4.5}$$

Applying far-field condition, i.e., $r \to \infty$, $\phi \to 0$, we get $c_2 = 0$, $c_1 = \dot{R}R^2$, and the velocity potential as

$$\phi = -\frac{\dot{R}R^2}{r} \tag{4.6}$$

So, the radial velocity is $u_r = \frac{\partial \phi}{\partial r} = \frac{\dot{R}R^2}{r^2}$, the tangential velocity is $u_\theta = 0$, and the resultant velocity is

$$v = \sqrt{\left(u_r^2 + u_\theta^2\right)} = \frac{\dot{R}R^2}{r^2} \tag{4.7}$$

The kinetic energy of the cavity $U_K$ is expressed (Engel 1967) as

$$U_K = \int \frac{\rho}{2} v^2 dV = \int_R^\infty \frac{\rho}{2} \frac{\dot{R}^2 R^4}{r^4} 2\pi r^2 dr = \pi \rho R^3 \dot{R}^2 \tag{4.8}$$

The kinetic energy of the impacting droplet is

$$E_K = \frac{2}{3}\pi R_o^3 \rho V_0^2 \tag{4.9}$$

where, $V_o$ is the impact velocity of the droplet, and $R_o$ is the droplet radius. To account for the kinetic energy of the cavity and the initial droplet impact, two correction factors, $\alpha$ and $\beta$ , are introduced. The $\alpha$ is applied because the cavity kinetic energy is estimated using a simplified velocity model. The $\beta$ compensates for energy losses that are neglected in the model, such as minor viscous dissipation, and the energy associated with crown formation. From the conservation of energy

$$U_P + U_S + \alpha U_K = \beta E_K \tag{4.10}$$

$$\frac{\pi g \rho R^4}{4} + \pi R^2 \sigma + \alpha \pi \rho R^3 \dot{R}^2 = \beta \frac{2}{3}\pi R_o^3 \rho V_o^2 \tag{4.11}$$

### 4.2.1 *Non-dimensionalization and solution of the energy equation*

We non-dimensionalize the energy equation with appropriate quantities as:

$R^* = \frac{R}{R_o}$, $t^* = \frac{tV_o}{R_o}$ and $\dot{R}^* = \frac{\dot{R}}{V_o}$. The corresponding dimensionless form of equation 4.11 is:

$$\frac{1}{4}\frac{gR_o}{V_o^2} R^{*4} + \frac{\sigma}{\rho R_0 V_0^2} R^{*2} + \alpha R^{*3} \dot{R}^{*2} = \frac{2}{3}\beta \tag{4.12}$$

$$\frac{1}{4}\frac{1}{Fr} R^{*4} + \frac{1}{We} R^{*2} + \alpha R^{*3} \dot{R}^{*2} = \frac{2}{3}\beta \tag{4.13}$$

At maximum expansion of the cavity (at the verge of retraction), $R^* = R^*_{\max}$ and $\dot{R}^* = 0$. Using these conditions, the factor $\beta$ is obtained from the experimental measurement of $R^*_{\max}$ as

$$\beta = \frac{3}{2}\left[\frac{1}{4}\frac{1}{Fr}(R^*_{\max})^4 + \frac{1}{We}(R^*_{\max})^2\right] \tag{4.14}$$

Once $\beta$ is determined, the parameter $\alpha$ is adjusted to achieve the best agreement between the model predictions and experimentally measured time evolution of the mean cavity radius.

We now solve the differential equation Eq. (4.13), imposing the initial boundary condition $R^*(1)=1$. This assumption implies that immediately after the impact, the cavity radius is equal to the radius of the impacting droplet at $t=\frac{R_o}{V_o}$ ($t^*=1$).

As the expanding cavity radius ($R$) is a function of $\theta$ and $t$, to determine the mean cavity radius, we appeal to shifted Legendre polynomials. The shifted Legendre polynomials are defined as an affine transformation of standard Legendre polynomials: $\tilde{P}_n(x)=P_n(2x-1)$, where the shifting function $x$ to ($2x-1$) maps the interval [-1,1] to [0,1]. The transformation implies that the polynomials $\tilde{P}_n(x)$ are orthogonal on [0,1]. We obtain the mean cavity radius from the $n=0$ coefficient, that is $R_{mean}=a_0$, while $n\geq 1$ retain orthogonality on transformed domain. The mean cavity radius is obtained as:

$R_\theta(\theta,t)=\sum_{n=0}^{5} a_n\tilde{P}_n(\cos\theta)$, where $a_n$ is an experimentally obtained coefficient, and we use Legendre polynomials $\tilde{P}_n$ up to degree $n=5$. For a particular time-instant, we extracted the cavity radius for several polar angles, $\theta$, using image processing, and obtained the experimental mean cavity radius, i.e., $a_0$.

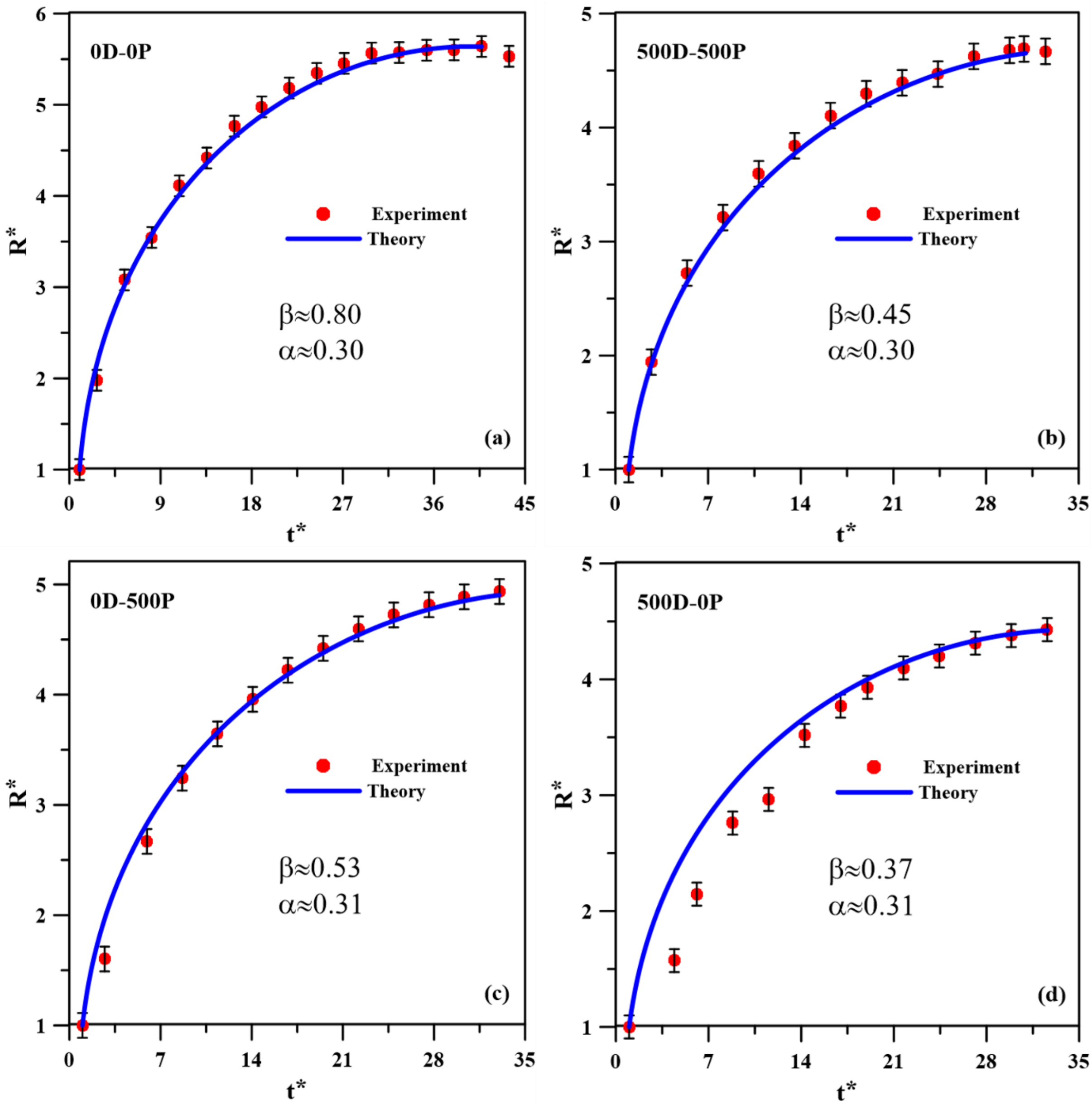

**Figure 10. Temporal evolution of the non-dimensional mean cavity radius $R^*$ for different droplet-pool combinations, at $We = 192$ and $Fr = 678$: (a) water droplet impacting its own pool, (b) 500 ppm elastic fluid droplet impacting its own pool, (c) water droplet impacting a 500 ppm elastic fluid pool, and (d) 500 ppm elastic fluid droplet impacting a water pool. The red circles represent the experimental data, and blue solid line corresponds to the energy conservation equation (Eq. (4.13)).**

Figure 10 shows the time-evolution of cavity radius different droplet-pool combinations. The graphs compare the experimental data with the best fit energy model (Eqn. 4.13), by tuning the factors. In all cases, careful and physically consistent use of the factors leads to close agreement between the experimental data and the model predictions. By comparing the value of $\beta$ corresponding to the 0D-0P case, with those of the other droplet-pool combinations, we quantify the amount of elastic energy stored by the stretching of the long-chain polymer molecules during cavity formation. In the 0D-0P combination, we obtain $\beta = 0.80$, i.e., ~80% of the initial kinetic energy of the impacting droplet contributes to the cavity expansion process. In contrast, for the 500D-500P combination, we obtain $\beta = 0.45$, i.e. only ~45% of the initial kinetic energy contributes to cavity expansion processes. This reduction indicates that ~35% of the initial kinetic energy is stored (0.8 in 0D-0P case and 0.45 in 500D-500P case) in the form of elastic energy by the polymer molecules (present in the system) stretching due to shear along with the cavity expansion process.

Now the question naturally arises, that since the absorbed energy comprises of elastic energy stored in the polymer molecules of both the impacting droplet and the target pool, can the relative contributions be mapped? To segregate the quantum of contribution of the polymer chains within the droplet, the intermediate configurations, 0D-500P and 500D-0P, are examined. For the 0D-500P case, we obtain $\beta = 0.53$, i.e. ~53% of the initial kinetic energy is utilized towards cavity expansion processes. Herein, since the impacting droplet is devoid of polymers, its contribution towards elastic energy storage is zero. Consequently, this ~ 0.27 value reduction in $\beta$ (0.8 for 0D-0P case and 0.53 in this case) is attributed entirely to elastic energy storage by the polymers present in the pool.

Likewise, for the 500D-0P case, we obtain $\beta = 0.37$, i.e. ~37% of the initial kinetic energy contributes to cavity expansion process. Here, the pool is devoid of any polymers, and hence the ~0.43 value reduction in $\beta$ (0.8 for 0D-0P case and 0.37 in this case) is due to the elastic energy stored exclusively by the droplet fluid. The maximum cavity radius is $R^*_{\max}$ =4.94 for 0D-500P case, and $R^*_{\max}$=4.4 for the 500D-0P case, indicating only a small difference between the two cases. However, $\beta$ is a function of ${R^*_{\max}}^4$. Consequently, even a small difference in the maximum cavity radius results in a relatively large change in $\beta$. Therefore, $\beta$ is 0.37 for 500D-0P cases. This apparent anomaly may be attributed to the elasticity of the droplet, which reduces the initial momentum transferred from the droplet to the pool compared with that of a pure water droplet. Consequently, for the 500D-0P combination, certain deviations between the experimental data and theoretically best-fit model are observed (figure 10d) during the early period because the cavity herein initially has a more inverted-bucket-like shape, rather than the hemispherical shape assumed in the model. At the later stages of cavity expansion, it assumes the hemispherical shape, leading to good agreement between the experimental results and the best-fit theory.

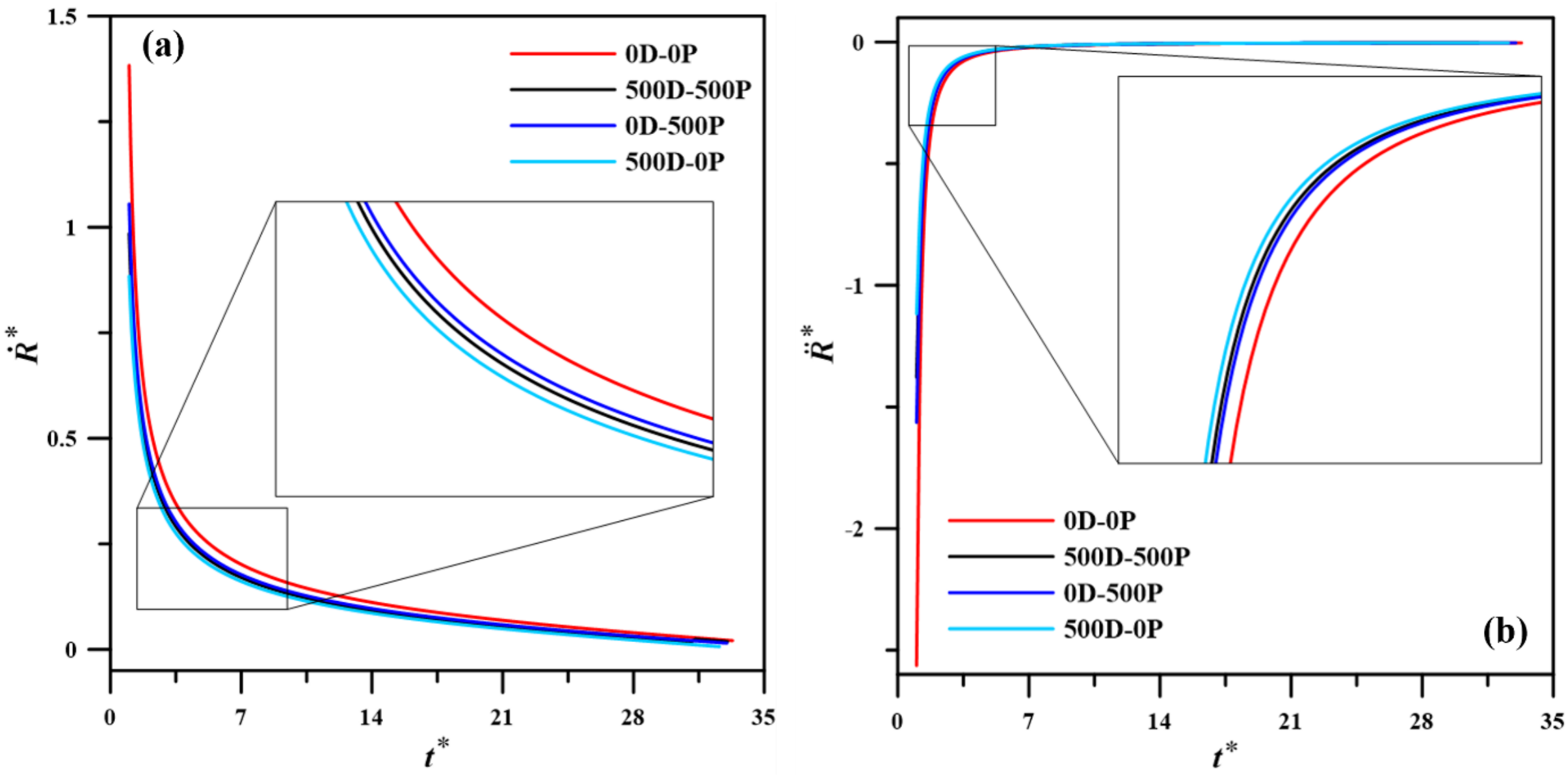

**Figure 11. (a) time-evolution of normalised mean cavity velocity for different combinations of droplet-pool interactions, and (b) time-evolution of mean cavity acceleration.**

Figure 11 illustrates the time-evolution of the velocity and acceleration of the cavity during expansion regime. At early time frames, the cavity growth velocity is very high (Figure 11a), and the potential and surface energies of the cavity are negligible compared to its kinetic energy. This implies that the kinetic energy of the impacting droplet is converted almost entirely into the kinetic energy of the flow surrounding the cavity. At later time frames, the cavity velocity dies out to small values (Figure 11a). During the early stages of expansion, the velocity is lower for impacts involving elastic liquids than for the pure water case. At these early times, the high cavity velocity generates large localized shear rates, causes the stretching of the polymer chains during expansion. Also, the shear rates are greater than the inverse of the polymer relaxation timescales. Under these conditions, the polymer chains do not have sufficient time to undergo stress relaxation, and therefore remain stretched. The stretching of the polymer chains store certain portion of the flow kinetic energy as elastic energy, thus reducing the kinetic energy available for cavity expansion, and consequently lowering the cavity propagation velocity compared to the Newtonian case. At later time frames, the cavity velocity decreases substantially, leading to a corresponding reduction in the local shear rates. This scenario permits the stretched polymer chains to relax, diminishing the influence of elastic stresses on the corresponding flow. Thus, the cavity velocity plots for the elastic and Newtonian cases converge during the later stages of cavity expansion. Due to this, the deceleration rates of the cavity growth process at later time frames are faster for elastic fluids than the Newtonian counterpart (see figure 11b).

### 4.3. *Influence of elastic stress on jet formation and evolution*

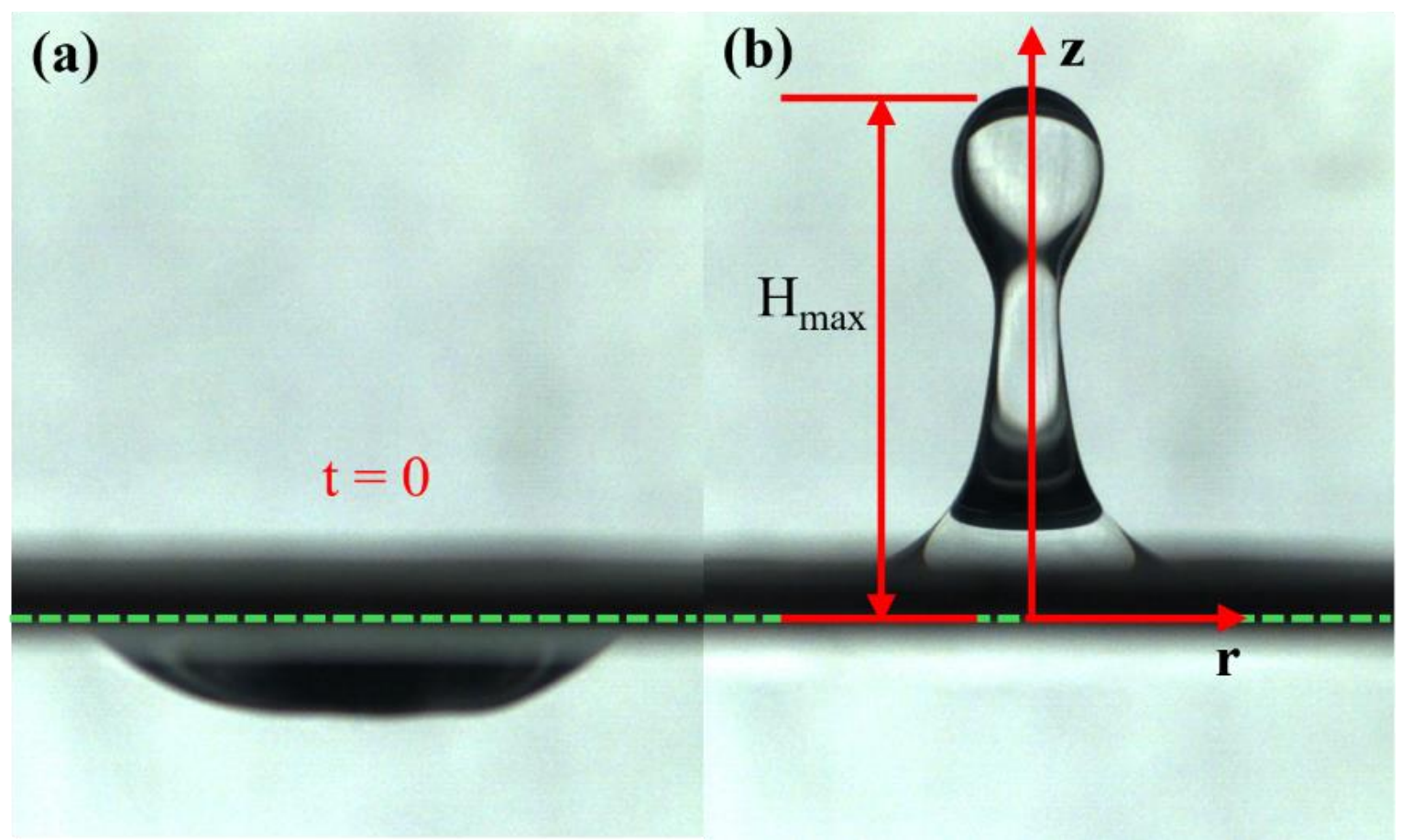

**Figure 12. (a) Experimental snapshot of $t = 0$, at which the jet first emerges following the collapse of the capillary wave at the base of the cavity, and (b) representation of the *r-z* coordinate system for the fully evolved Worthington jet (at maximum jet height).**

We analyze the spatio-temporal evolution of the elastic-liquid Worthington jet, for $Oh \ll 1$ and $De \ll 1$. The collapse of the capillary wave at the crater base leads to the formation of the Worthington jet. The jet is then driven by inertia, while capillarity and elasticity resist the same. We ascertain the jet dynamics from the temporal evolution of the jet height, H, and the minimum jet radius $r_{\min}$. The variation of normalized jet height $H^*$ with normalized time $\tilde{t}$ (experimental) is shown in figure 12, for $We = 192$ and different $De$. Here, $De$ varies in as 0.01-0.05, and $Oh$ varies as 0.0031-0.0046. In the following discussion, all heights are normalized by the droplet radius $R_o$, and time is normalized by the inertio-capillary time $t_c = \sqrt{\dfrac{\rho R_o^3}{\sigma}}$. Although normalizing the jet height by the jet radius would be physically intuitive, the jet radius is not constant and varies along the length of the jet. Thus, to ensure a consistent and well-defined scaling, the $R_o$ is used. Here, $t = 0$ (or $\tilde{t}$ =0) represents the time instant when the jet first emerges above the cavity base during retraction phase (see figure 12a). Since the jet remains sparsely observable until it rises above the initial undisturbed free surface; the initial jet height is not presented in the graph.

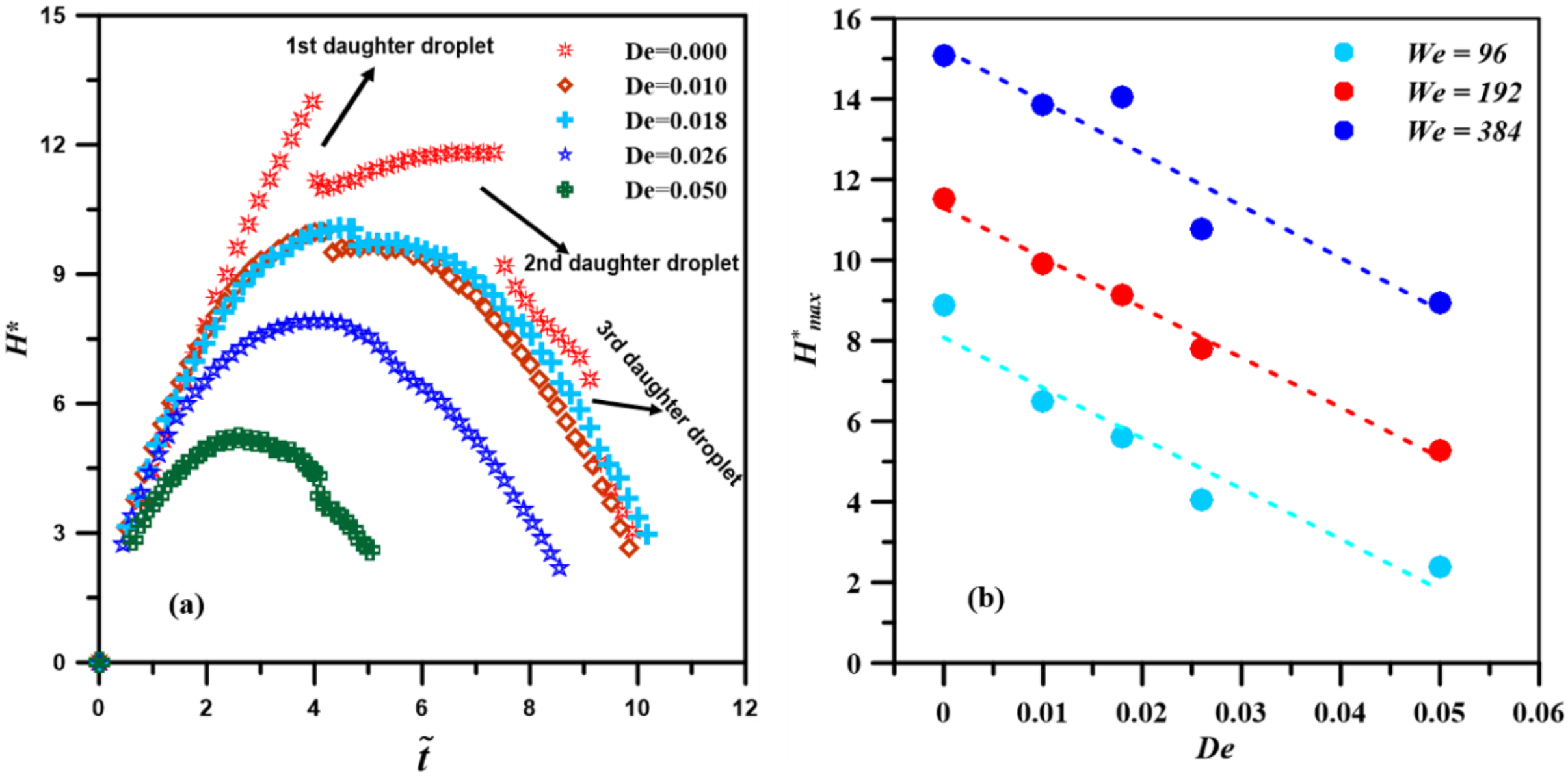

**Figure 13. (a) Temporal evolution of jet height for varying $De$ at $We = 192$, (b) effect of elastic stress relaxation on the maximum height of Worthington jet. The droplet and the pool liquid contain same polymer concentration.**

Figure 13a shows the temporal evolution of jet height, for different $De$, at $We = 192$. The central Worthington jet emerges from the base of the collapsing cavity as the capillary waves converge along the axis of symmetry. Post-birth, the jet height increases nonlinearly until it a maxima, after which the jet collapses back under gravity. For the Newtonian case ( $De$=0), the jet achieves the maximum height among all other cases, as there is no elastic stress to oppose the inertia. As the *De* increases, the polymer chains within the developing jet undergo progressive stretching, generating elastic stresses that resist the upward motion of the jet. Consequently, the maximum jet height decreases. It can be observed from figure 13a that, for the Newtonian case, the jet height increases linearly for a significant amount of initial period. As the *De* increases, the jet growth follows a sub-linear trend owing to slower relaxation of elastic stress.

During the early stages of jet growth ($\tilde{t} < 1$), all the curves collapse onto a single locus, indicating that elastic stresses have a negligible influence at the instant when the jet emerges, and is purely inertia driven. For $\tilde{t} > 1$, however, elastic stresses begin to influence the jet evolution, leading to deviations from the Newtonian behaviour, wherein only capillary stresses are present. Although polymer relaxation delays the propagation of the capillary waves along the cavity, it does not affect the initial growth of the jet. This observation suggests that at the instant of jet formation, when the capillary waves converge along the axis of symmetry, the polymer molecules recoil back to their equilibrium (un-stretched) configuration, and therefore generate negligible elastic stress. Figure 13b illustrates the dependence of the non-dimensional maximum jet height on $De$, for three different $We$ (=96, 192, and 384). The best-fit lines in figure 13b indicate that the trend with which the maximum height of the jet varies is nearly-linear with respect to $De$. For a fixed $We$, the maximum jet height decreases linearly with increasing $De$ due to higher degree of energy dissipation via polymer stretching and slower elastic stress relaxation, which eventually decelerates the growth of the jet.

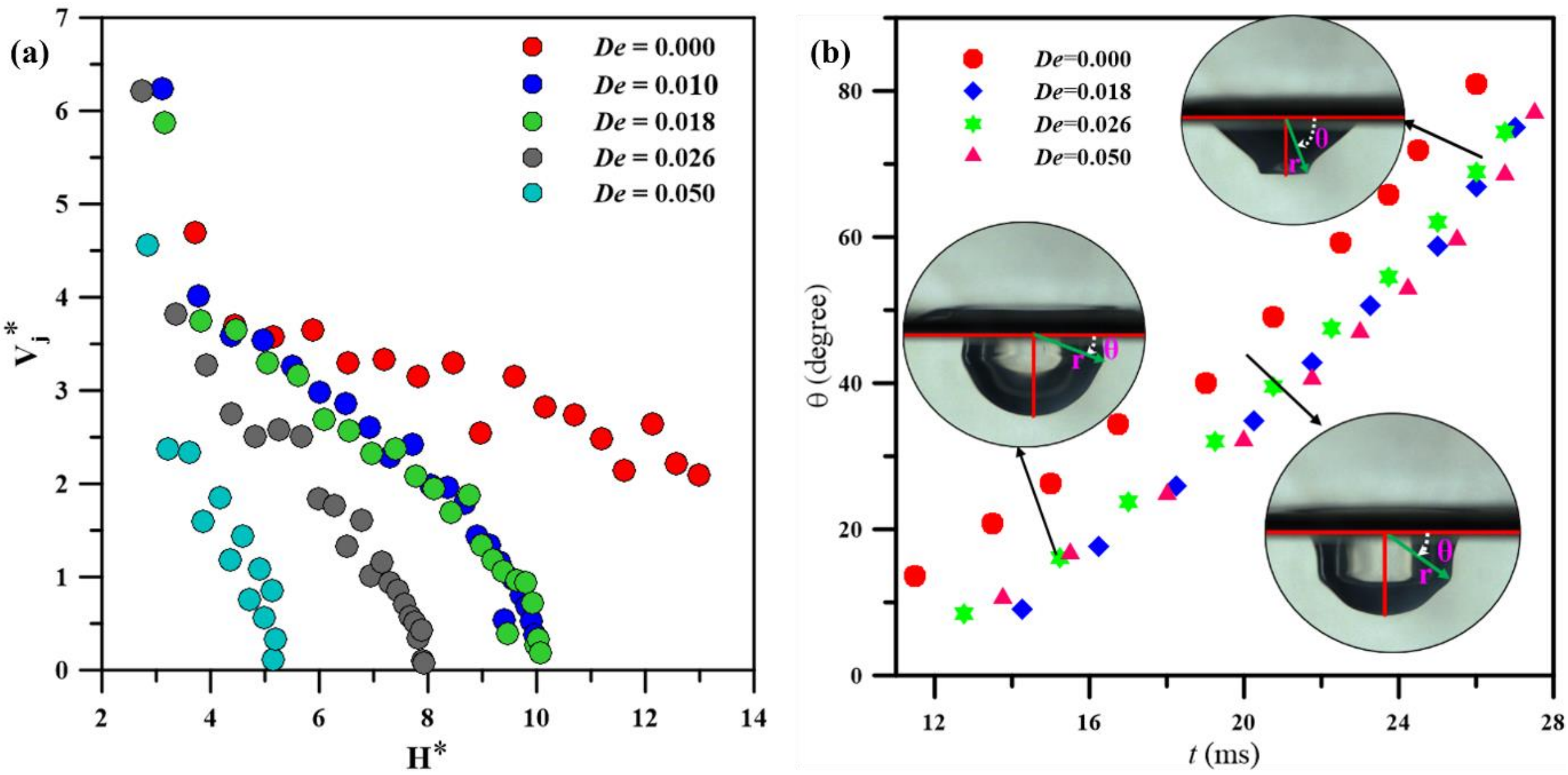

**Figure 14. (a) Variation of non-dimensional jet-tip velocity with non-dimensional jet height, for different $De$, at $We$=192, (b) trajectory of capillary wave propagation along the cavity walls for different $De$. The angular position of the capillary wave was measured with respect to time, in a clockwise sense from the undisturbed free surface ($\theta = 0$). The droplet and the pool consist of the same liquid in both figures.**

Figure 14a presents the variation of the jet-tip velocity with jet height up to its maximum rise. The jet-tip velocity is normalized by the capillary velocity scale, $V_c = \sqrt{\frac{\sigma}{\rho R_o}}$.

As the $De$ increases, velocity-height curves become steeper (figure 14a), indicating that the jet-tip velocity decays at a faster rate with increasing jet height. However, the initial jet-tip ejection velocity remains largely unchanged with $De$. As discussed earlier, during the convergence of the capillary waves, the long polymer chains recoil to their initial equilibrium configuration, making the initial jet ejection velocity independent of $De$. From figure 14b, it can be observed that for $De$=0, the capillary waves propagate faster than in the elastic fluids. As a result, the onset of jet ejection is delayed with increasing $De$; however, the initial jet ejection velocity remains nearly the same for all cases. As the jet rises, both capillary and elastic forces oppose its upward motion. Therefore, with increasing $De$, the jet-tip velocity decreases at a faster rate due to the slower relaxation of elastic stresses and the greater conversion of the jet's kinetic energy into elastic energy. In contrast, for the Newtonian case ( $De$=0), no elastic stresses are present to resist the jet motion, causing the jet velocity to decrease at a slower rate.

**4.4.** ***Elasto-hydrodynamics of the stretched elasto-capillary Worthington jet***

Next, we analyse the thinning, and the pinch-off behaviour of the stretched, elasto-capillary Worthington jet to unveil the effects of fluid elasticity. The thin filaments connecting the satellite droplet with primary jet body develop due to the capillary instability and the stretching of polymer chains. We assume the elongated Worthington jet as an elastic jet under uniaxial tension. It is evident from figure 16b that the jet is in the elasto-capillary regime for $t > 40$ ms, where the capillary pressure and elastic stresses jointly govern the jet dynamics. For our case, a similar (0+1) dimensional theory, developed for the capillary thinning of viscoelastic fluids in the Capillary Breakup Extensional Rheometer (Entov & Hinch 1997; Bazilevsky *et al.* 1990; Rozhkov 1983), is applied with the incorporation of a axial tensile force that remains constant along the length of the jet. The net tensile force $T$, acting along the jet at any cross-section (Clasen *et al.* 2009) reads

$$T = \pi r \sigma + \pi r^2 \tau_e \qquad (4.15)$$

Where, $\tau_e$ is the elastic stress generated due to the stretching of the dilute, solvated polymer chains. At $t = 0$, when the jet emerges above due to the capillary wave collapse at the base of the cavity, we assume that the polymers recoil to their equilibrium initial configuration, providing negligible elastic stress. Thus, initially

$$T = \pi r_o \sigma \qquad (4.16)$$

Where, $r_o$ is the jet radius at $t=0$. As the tensile force is constant on any cross-section along the jet, equating 4.15 and 4.16, we get the elastic stress contribution as

$$\tau_e = \frac{\sigma}{r}\left(\frac{r_o}{r}-1\right) \tag{4.17}$$

For the *satellite droplet + filament* regime in our experiments of elastic fluids, $Oh = 0.0046$, $We = 386$, and $De = 0.01\text{-}0.05$. Thus, the viscous stresses are neglected, while elastic stress contribution is prominent due to large stretching strain rate.

We appeal to the FENE-P model (Bird et al. 1987) to analyze the thinning dynamics of the elastic Worthington jet. This model incorporates the finite extensibility of the polymer molecules; and this constitutive non-linear elastic model has previously been applied successfully to examine the thinning dynamics of slender viscoelastic filaments (Entov & Hinch 1997; Clasen *et al.* 2006). The components of the elastic deformation tensor $\boldsymbol{A}$ in the FENE model, considering single relaxation mode, are expressed as:

$$\frac{dA_{zz}}{dt} = 2E_{zz}A_{zz} - \frac{f}{\lambda}(A_{zz}-1) \tag{4.18}$$

$$\frac{dA_{rr}}{dt} = 2E_{rr}A_{rr} - \frac{f}{\lambda}(A_{rr}-1) \tag{4.19}$$

Here, $E_{zz}$ and $E_{rr}$ are the axial and radial components of the velocity gradient (the strain rate tensor), $\lambda$ is the relaxation time, and $f$ is a correction term accounting for the finite extensibility of the polymer molecules, expressed as

$$f = \frac{1}{(1+\mathrm{tr}(\mathbf{I})/b) - (A_{zz}+2A_{rr})/b} \tag{4.20}$$

Here, $\mathbf{I}$ is the unit tensor, and the extensibility parameter $b$ being the limit of $A_{zz}$ at the maximum extension of the polymer chains. For a cylindrical filament, the axial and radial components of the strain rate tensor are (Entov & Hinch 1997):

$$E_{zz} = -\frac{2\dot{r}}{r},\ E_{rr} = \frac{\dot{r}}{r} \tag{4.21}$$

The components of the viscoelastic stress tensor within FENE model are expressed as

$$\sigma_{zz} = 2\mu E_{zz} + Gf(A_{zz} - 1) \quad (4.22)$$

$$\sigma_{rr} = 2\mu E_{rr} + Gf(A_{rr} - 1) \quad (4.23)$$

Now subtracting Eq. (4.23) from Eq. (4.22), we obtain

$$\sigma_{zz} - \sigma_{rr} = 2\mu(E_{zz} - E_{rr}) + Gf(A_{zz} - A_{rr}) \quad (4.24)$$

The elastic stress $\tau_e$ in Eq. (4.17) is then given by

$$\tau_e = 2\mu(E_{zz} - E_{rr}) + Gf(A_{zz} - A_{rr}) \quad (4.25)$$

Incorporating Eq. (4.21) in Eq. (4.25) yields

$$\tau_e = -6\mu\left(\frac{\dot{r}}{r}\right) + Gf(A_{zz} - A_{rr}) \quad (4.26)$$

Now equating Eq. (4.17), and Eq. (4.26), we get

$$6\mu\left(\frac{\dot{r}}{r}\right) = -\frac{\sigma r_o}{r^2} + \frac{\sigma}{r} + Gf(A_{zz} - A_{rr}) \quad (4.27)$$

Substituting $E_{zz}$ and $E_{rr}$ in Eq. (4.18), and (4.19), the components of the deformation tensor take the following form:

$$\frac{dA_{zz}}{dt} + 4\left(\frac{\dot{r}}{r}\right)A_{zz} = -\frac{f}{\lambda}(A_{zz} - 1) \quad (4.28)$$

$$\frac{dA_{rr}}{dt} - 2\left(\frac{\dot{r}}{r}\right)A_{rr} = -\frac{f}{\lambda}(A_{rr} - 1) \quad (4.29)$$

In our study, we focus on the elasto-capillary regime, where axial deformation in the slender elastic jet becomes very large compared to radial deformation, i.e. $A_{zz} \gg 1; A_{zz} \gg A_{rr}$. We also assume that the polymer molecules are highly extensible in nature and do not reach their finite extensibility limit during stretching under the studied shear rates. Under this condition: $A_{zz} \ll b,\ f \approx 1$. Therefore, Eq. (4.27) and (4.28) can be reduced to

$$6\mu\left(\frac{\dot{r}}{r}\right) = -\frac{\sigma r_o}{r^2} + \frac{\sigma}{r} + GA_{zz} \quad (4.30)$$

$$\frac{dA_{zz}}{dt} + 4\left(\frac{\dot{r}}{r}\right)A_{zz} = -\frac{1}{\lambda}A_{zz} \tag{4.31}$$

Under the assumption that the polymers recoil to their initial configuration at the base of the jet, integration of Eq. (4.31) leads to the relation

$$A_{zz}r^4 = A_{zz}\big|_{z=0} r_o^{\,4} \tag{4.32}$$

Where $r_o$ represents the jet radius at $t$=0 and $z$=0. By definition, when the polymers are in their recoiled configuration (un-stretched), the conformation tensor is equal to the unit tensor, i.e., $\mathbf{A} = \mathbf{I}$, $A_{zz}\big|_{z=0,t=0} = 1$. Now, using this initial condition, Eq. (4.31) can be solved together with Eq. (4.32), yielding an exponential decay of $A_{zz}$

$$A_{zz}r^4 = r_o^4 e^{-t/\lambda} \tag{4.33}$$

Substituting Eq. (4.33) into Eq. (4.30) leads to an approximate governing equation for the temporal evolution of the jet radius, as

$$6\mu\left(\frac{\dot{r}}{r}\right) = -\frac{\sigma r_o}{r^2} + \frac{\sigma}{r} + \frac{G r_o^{\,4}}{r^4} e^{-t/\lambda} \tag{4.34}$$

Where, by definition, $G = \frac{\mu_p}{\lambda}$. In the elasto-capillary regime, as $r$ becomes very small, the dynamics are primarily governed by a balance between the $\frac{\sigma r_o}{r^2}$ and $\frac{G r_o^{\,4}}{r^4} e^{-t/\lambda}$ terms appearing in Eq. (4.34). Hence, the solution exhibits exponential asymptotic behaviour in the long-time limit (Eqn. 4.35), which is also well corroborated by the experimental data and scaling curve shown in Figure 17.

$$\frac{r}{r_o} \sim \sqrt{\frac{G r_o}{\sigma}} e^{-t/2\lambda} \tag{4.35}$$

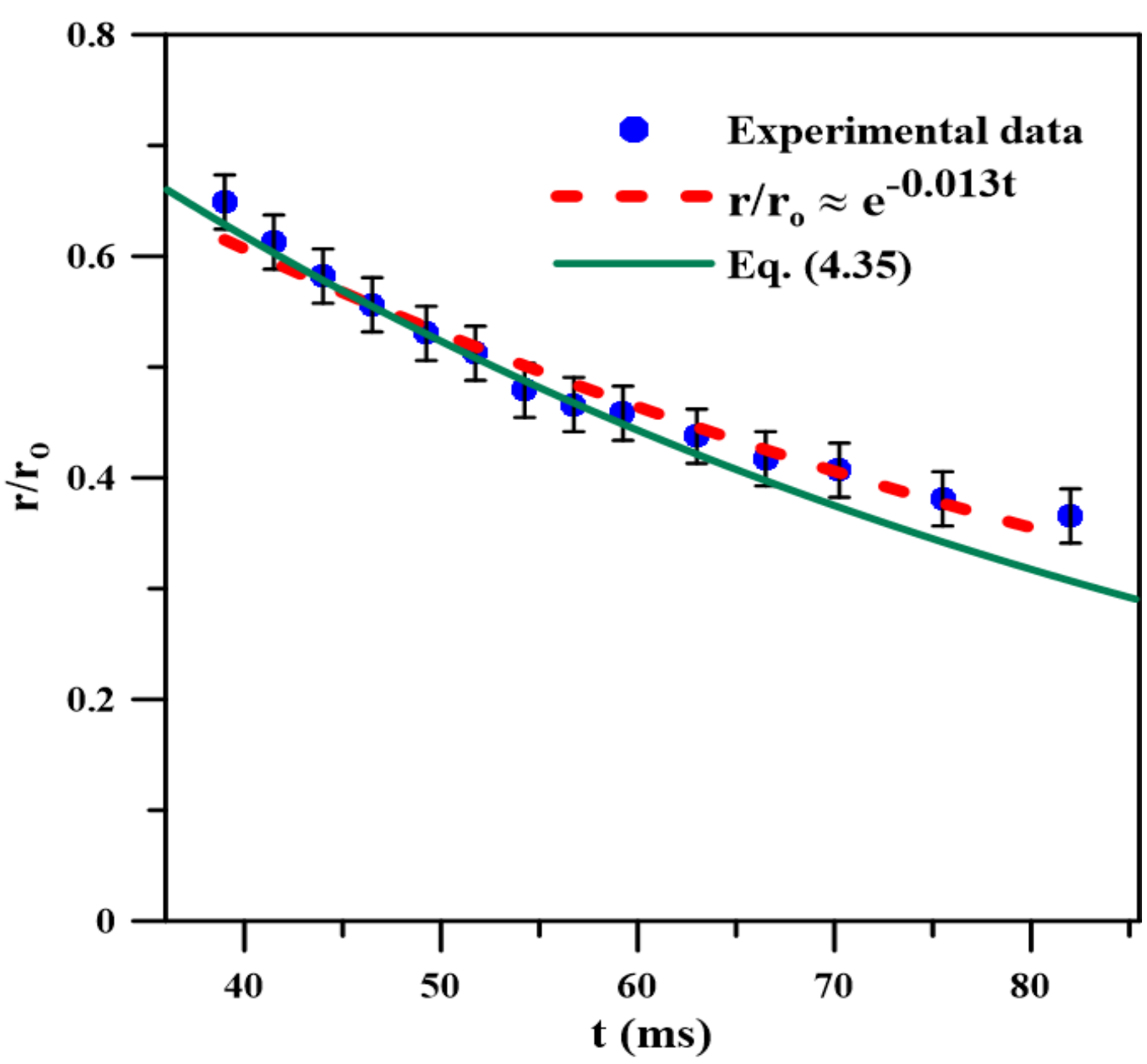

**Figure 15. Comparison between the experimentally measured thinning of the dimensionless elastic Worthington jet radius; and the theoretical prediction from the one-dimensional elasto-capillary thinning model (Eq. 4.35). The comparison is shown here for the 500D-500P case, at *We* = 384.**

Hence, we infer that the dynamics of the elastic Worthington jet observed in *satellite droplet + filament* regime are governed by Eq. (4.34), presenting a balance between elastic and capillary forces. Interestingly, Eq. (4.35) further reveals that the radius of the Worthington jet in the elasto-capillary regime decreases exponentially (see figure 15) as $r(t) \sim \exp(-t/2\lambda)$, while the classical viscoelastic slender cylindrical filament investigated by Entov & Hinch (1997), and later works follows a $r(t) \sim \exp(-t/3\lambda)$ decay law. It is noteworthy that Eq. (4.35) overestimates the relaxation time by approximately two orders of magnitude, and similar discrepancies have also been reported by several previous studies (Doyle *et al.* 1998; Lindner *et al.* 2003) on viscoelastic filament thinning. The relaxation time predicted by Eq. (4.35) is approximately 29 ms, whereas the experimentally measured relaxation time of a 500 ppm aqueous PEO solution, obtained from extensional rheometry, is ~0.14 ms (Sen *et al.* 2022). Such mismatch might result from either polydispersity, or multiple relaxation time scales present in a single polymer chain under different conformation.

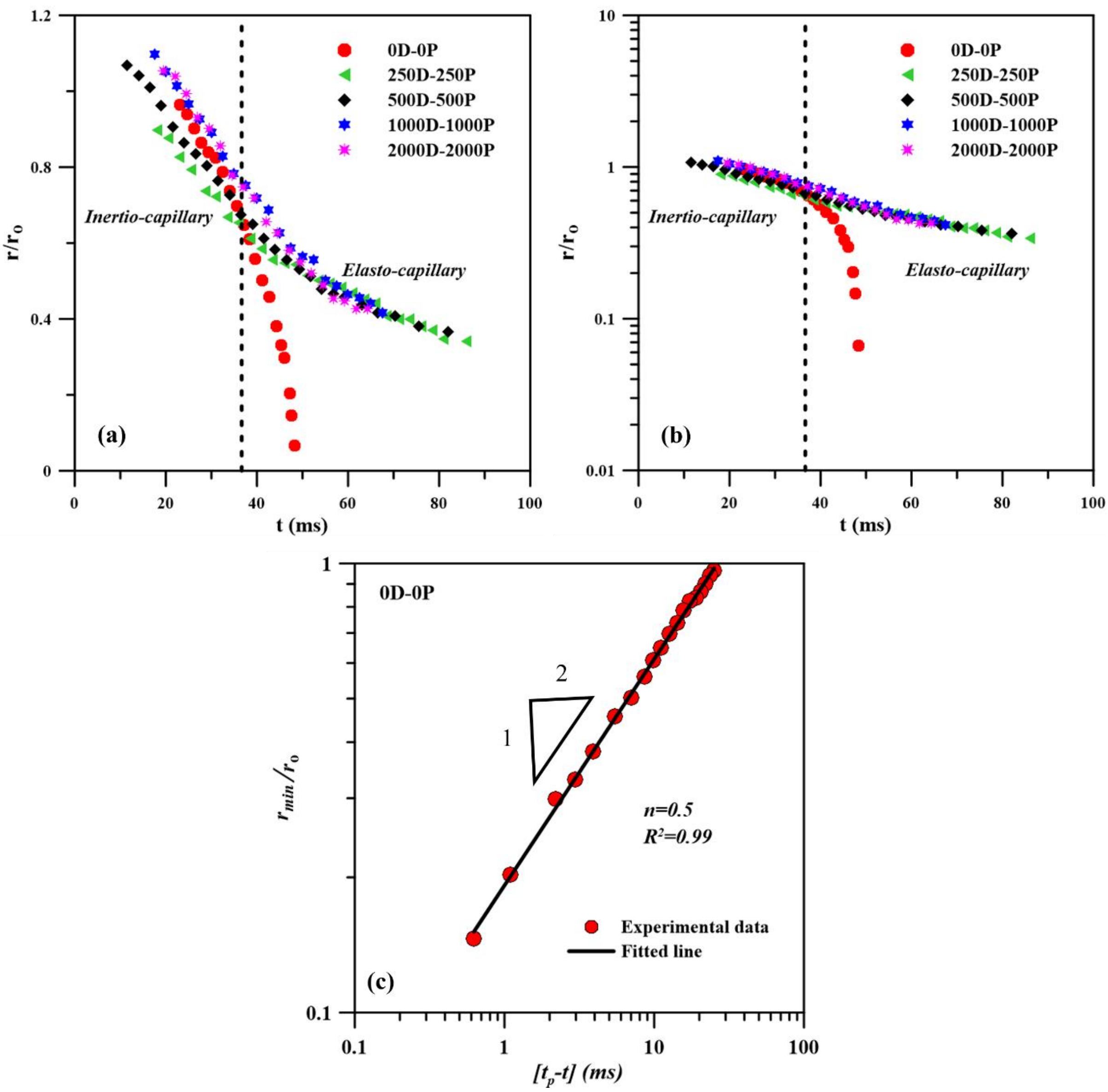

**Figure 16. Evolution of dimensionless jet radius, obtained from experimental snapshots with time, for different elastic fluid cases: (a) linear scale, (b) semi-logarithmic scale and (c) power law fitting of the Newtonian Worthington jet necking. For all experiments here, $Bo = 0.27$ and $We = 384$.**

In figure 16a, we have plotted the evolution of the Worthington jet thinning process as it approaches the pinch-off time, for the cases involving pure water (Newtonian case) and different elastic fluids. For the Newtonian case, the jet thinning follows a power-law scaling of $r_{\min}(t) \propto (t_P - t)^n$ (see figure 16c), where the exponent is $n = 0.5$, instead of the classical value of 2/3 reported in previous studies (Amarouchene *et al.* 2001; Cooper-White *et al.* 2002; Ambravaneswaran *et al.* 2002). The classical 2/3 scaling is derived by considering a balance solely between inertial and capillary forces during pinch-off. In contrast, the necking dynamics of the Worthington jet are influenced not only by inertia and surface tension, but also by gravity, which continuously decelerates the upward-rising jet. Consequently, although the neck radius still exhibits power-law behavior as the pinch-off time approaches, the scaling exponent differs from the classical inertio-capillary value due to the additional influence of gravity. At time $t > 40\,\mathrm{ms}$ (figure 16b), the filament radius of the elastic Worthington jet starts decreasing exponentially with time as $r(t) \sim \exp(-t/2\lambda)$, in agreement with the elasto-capillary thinning model. Figure 16b further shows that the filament thinning rate is nearly independent of the PEO concentration in the elastic fluids. Under the same impact velocity, neck deformation is reduced with increasing elasticity, thereby resulting in larger filament radius. As the elasticity increases, capillary thinning does not progress to a sufficiently advanced stage where differences in the thinning rate become apparent. If the impact velocity is increased further, leading to greater neck deformation and the formation of a thinner filament, different capillary thinning rates, corresponding to different levels of elasticity, may be observed in the long-time limit.

## 5. Numerical simulations

Since the distribution of elastic stresses cannot be determined experimentally, we perform computer simulations of a pure water case (*De*=0) and an elastic case (*De*=0.026) to elucidate the role of elastic stresses in the cavity and the jet dynamics. The simulations reveal the elastic stress distribution within the jet and demonstrate how elasticity alters the velocity field relative to the pure water case. Figure 17 illustrates a qualitative and quantitative comparison of the predictions from our simulations with respect to our experimental observations. We portray here an array, at different time-stamps, and try to validate the computational model over a wide range of regimes observed during the droplet-pool interaction events. The simulations show excellent agreement with the experiments, over all the stages of cavity formation and its collapse, and jet evolution. The cavity profiles, its depth and width, the formation and morphology of the crown, the jet structure and height, and the satellite droplet dimensions at different time instants are compared in figure 17. The close morphological and quantitative correspondence between the simulations and experimental results demonstrates the capability of the model to accurately capture the post-impact dynamics of elastic fluid droplet-pool interactions.

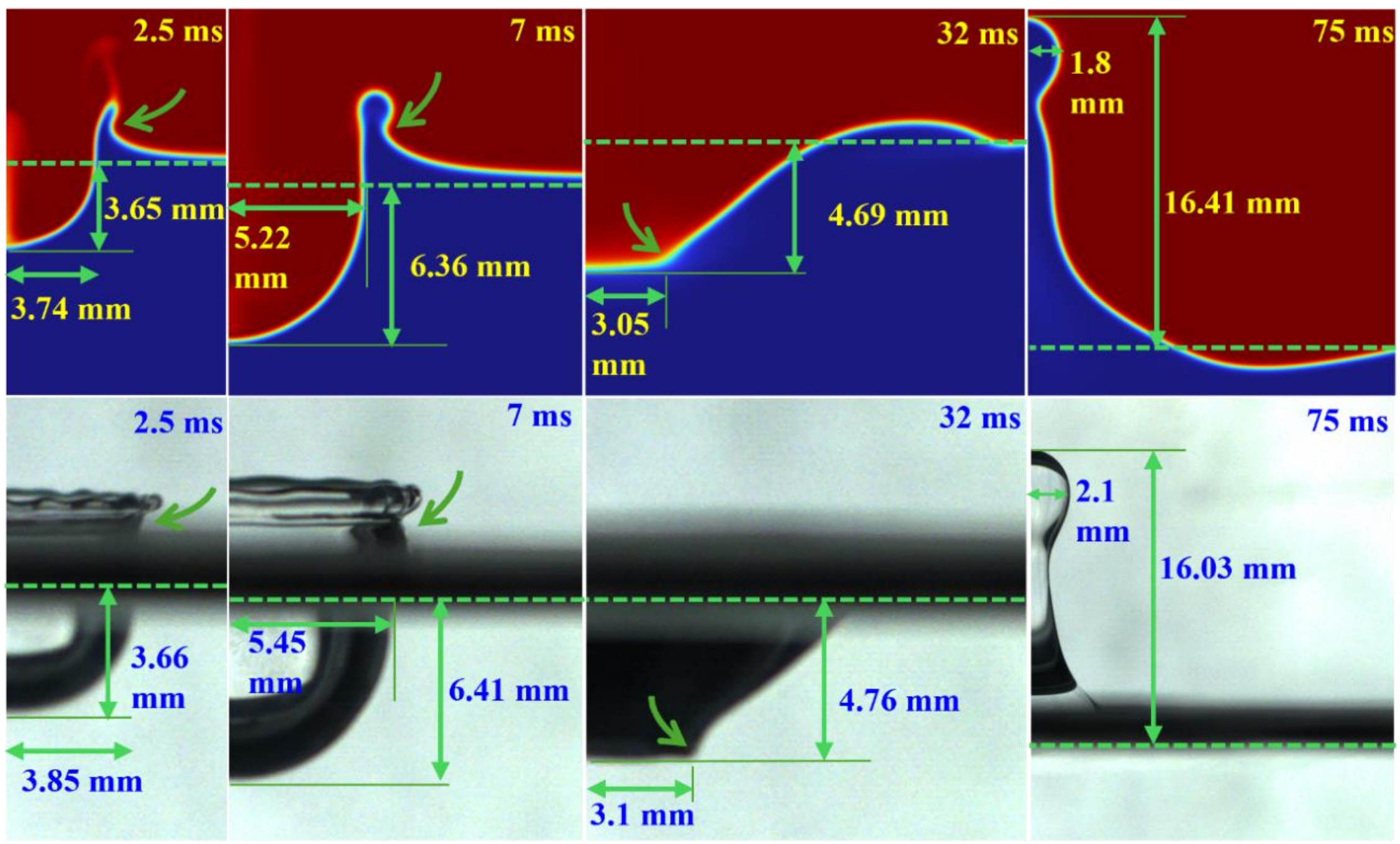

**Figure 17. Qualitative and quantitative comparison of events between the simulated findings and the experimental observations, during the impact of a 1000 ppm elastic fluid droplet on a 1000 ppm elastic fluid pool, at *We*=384, *Fr*=678, and $De = 0.026$. The simulation parameters are: $\lambda = 0.12$ ms, $\mu_s = 1$ mPa.s, $\mu_p = 0.37$ mPa.s, $\sigma = 0.06$ N m$^{-1}$,** $b = 2.5\times10^4$ ( Wagner *et al.* 2005)**. This figure serves as a validation for the computational model. At each time-snap, appreciable match between the simulations and experiments are noted, including several atypical morphologies (shown by curved arrows). The dotted horizontal lines represent the initial free surface of the pool.**

Figure 18 shows the evolution of zz-component of elastic stress tensor ($T_{p,zz}$) across the neck radius (the section indicated by the yellow arrow in inset) during the thinning process. As the neck radius reduces with time, the magnitude of $T_{p,zz}$ increases. The elastic stress is maximum at the centre of the neck (at the axis of symmetry) and decreases towards the free interface (figure 18a). To explain this radial variation in elastic stress, the axial velocity ($w$) at the same cross-section is plotted in figure 18b. The axial velocity is maximum at the centre (along the axis of symmetry) and decreases towards the liquid-air interface. This implies that the polymer molecules located near the axis of symmetry experience greater extensional stretching than those near the free surface. This enhanced stretching generates larger elastic stresses at the centre of the neck, whereas the weaker stretching near the interface results in lower elastic stresses.  Thus, the polymers present in the central core of the jet are primarily responsible for generating the elastic stress and contributing to the formation of the final thread. It can also be observed from Figure 18b that as the neck becomes thinner, the magnitude of the axial velocity increases. This is because the increasing curvature of the neck enhances the capillary pressure, which drives the liquid away from the neck region at a higher velocity.

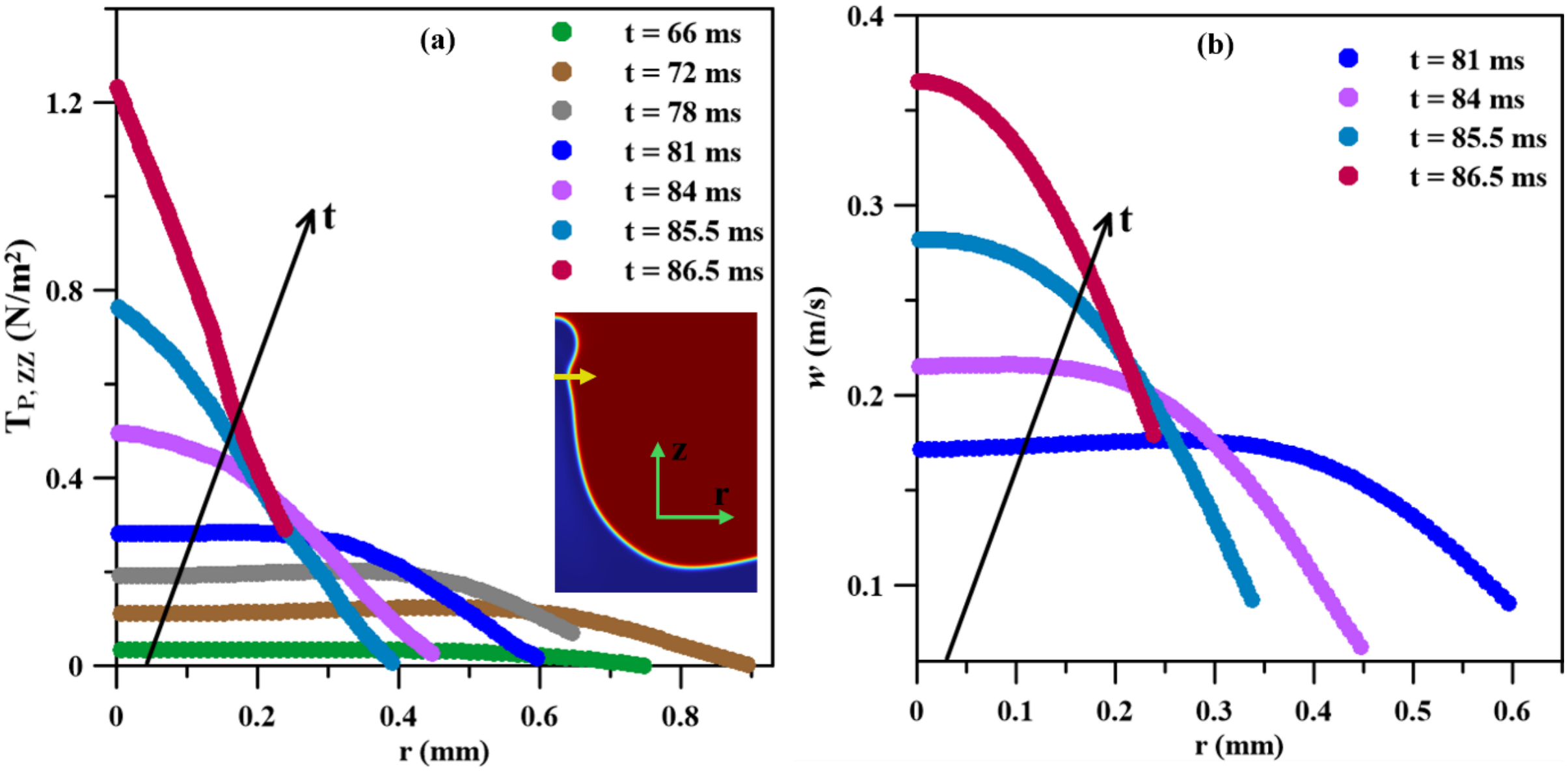

**Figure 18. (a) Evolution of the zz-component ($T_{p,zz}$) of the elastic stress tensor across a cross-section of the elastic Worthington jet at the neck region (indicated by the yellow arrow in the inset snapshot). (b) Variation of the axial velocity across the same cross-section.**

Figure 19 illustrates the evolution of the velocity field at different stages of cavity development and jet dynamics. At 2.5 ms after impact, as the cavity forms, neighboring air rushes into the cavity. A localized region (indicated by the dark red zone in the velocity contour) exhibits air velocities exceeding the droplet impact velocity (4.3 m/s), owing to the rapid creation of a low-pressure region within the expanding cavity, and the resulting large pressure gradient in the neighbourhood. From the velocity scale (colour bar) presented in figure 19 for the pure water (De=0.0) and elastic (De=0.026) cases, it is evident that, during the early stage of cavity expansion (2.5 ms), the cavity propagates at approximately twice the velocity in the pure water case compared with the elastic case. Hence, the air is entrained into the cavity at a significantly higher velocity (approximately two times higher than elastic case) in the pure water case. In the elastic case, the radial deformation beneath the impact point stretches the polymer molecules, generating elastic stresses that oppose the expansion of the cavity. As a result, the cavity propagates more slowly, thereby weakening the entrainment of air into the cavity and leading to a substantially lower air velocity.

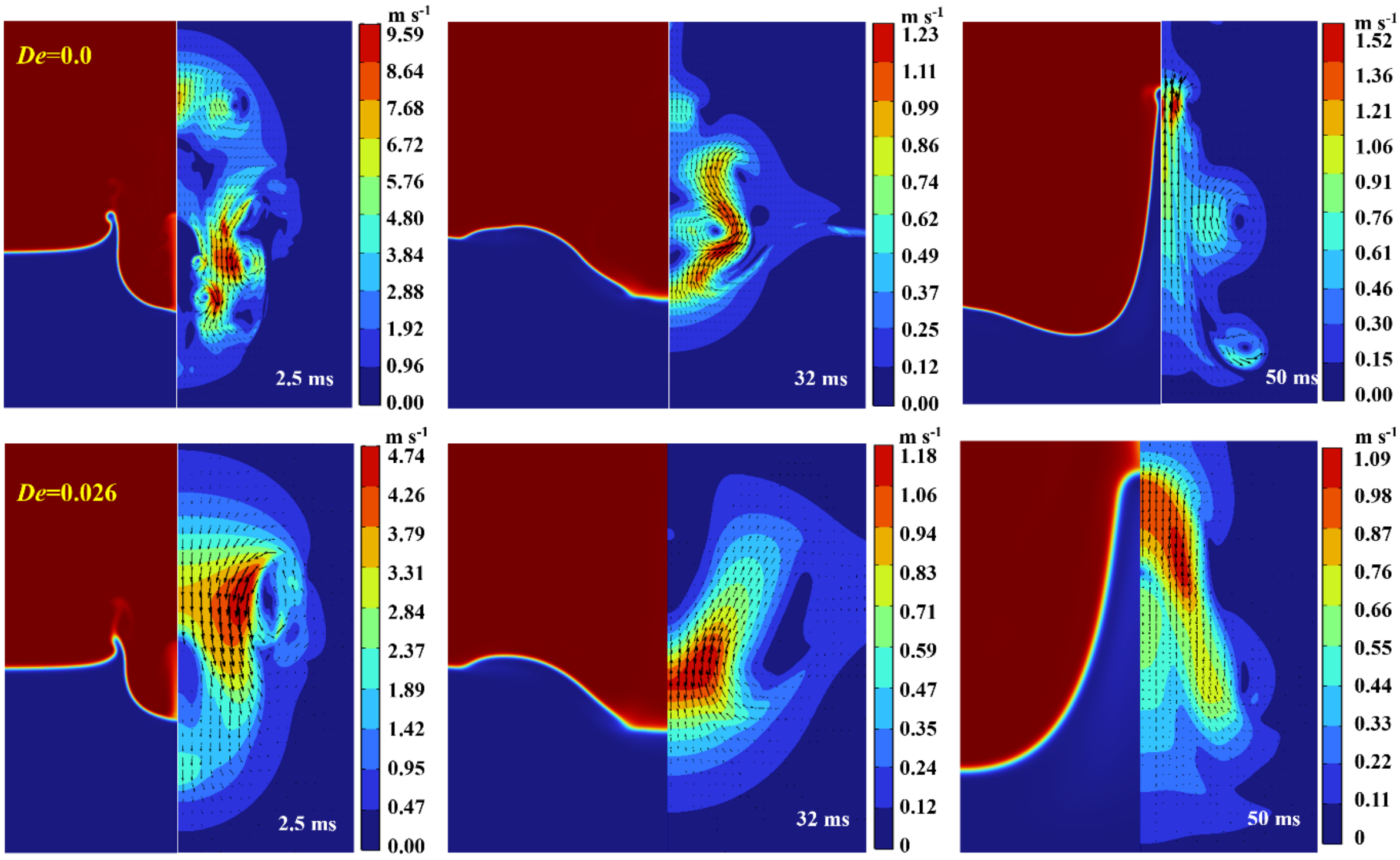

**Figure 19. Velocity contours and vectors at different instants during cavity and jet evolution. The numerical simulations reproduce the impact of a water droplet on a water pool (*De*=0) and an elastic droplet on an elastic pool (*De*=0.026) at $We = 384$, *Fr*=678. The simulation parameters are: $\lambda = 0.12$ ms, $\mu_s = 1$ mPa.s, $\mu_p = 0.37$ mPa.s, $\sigma = 0.06$ N m$^{-1}$, and $b = 2.5\times10^4$ ( Wagner *et al.* 2005).**

At t=32 ms, only a small difference is observed in the velocity magnitude between the two cases. This is because, during cavity reversal, the polymer molecules begin to recoil toward their equilibrium configuration and progressively relax, leading to a reduction in elastic stresses. An anticlockwise vortex is also observed inside the cavity during cavity reversal in the pure water case. However, no such vortex is observed in the elastic case, as the residual elastic stresses reduce the cavity retraction velocity compared to the pure water case. At t=50 ms, the fluid velocity within the jet is lower in the elastic case than in the pure water case because part of the fluid's kinetic energy is stored as elastic energy through polymer stretching. A weak, anticlockwise vortex is also observed near the crown, which promotes eventual crown breakup by shear. At 32 ms, during cavity retraction, the air within the cavity is expelled upward; however, its velocity is significantly lower than that observed during the early stage of cavity expansion (2.5 ms). In the liquid phase, the regions of high velocity are confined to the immediate vicinity of the cavity floor, while the velocity decays rapidly with increasing distance from the cavity. This observation supports the assumption of Prosperetti & Oguz 1993; that the velocity field vanishes simultaneously in the pool. From the velocity contours of the upward-moving jet at 50 ms, it is evident that the velocity at the jet tip is higher than that near the jet base, which corroborates with our experiments.

## 6. Conclusions

In our research, we investigate the elasto-hydrodynamics of elastic fluid droplets impacting onto a pool of elastic fluid, using detailed experiments, theoretical analysis and computer simulations. Our objective, and motivation has been to understand the role of fluid elasticity (in terms of elastic energy storage and polymer relaxation), capillarity and inertia in the fluid dynamic processes of cavity formation and collapse, crown formation, and growth and breakup of the Worthington jet, and the associated filament elasto-hydrodynamics. In particular, we demonstrate that the conversion of the droplet's initial kinetic energy into elastic energy, manifested via polymer chain stretching, significantly alters the impact dynamics by suppressing crown breakup, reducing cavity size and growth rates, and delaying jet evolution. We have observed distinct events, depending on the various droplet-pool combinations (water droplet-water pool, elastic fluid droplet-water pool, water droplet-elastic fluid pool, and elastic fluid droplet- elastic fluid pool), and governing $De$ and $We$, yielding four distinct regimes.

We put forward a theoretical estimation of the percentage of the initial kinetic energy of droplet that is converted into elastic energy during impact. As the polymer concentration increases, more than ~30% of the initial kinetic energy may be stored as elastic energy via polymer chain stretching during cavity expansion phase. This conversion of kinetic energy into elastic energy suppresses crown breakup, and the formation of secondary droplets. For a given $We$, the cavity size decreases with increasing polymer concentration owing to the enhanced conversion of kinetic energy into elastic energy. In particular, we obtain that the thinning of stretched Worthington jet radius follows the scaling $r(t) \sim \exp(-t/2\lambda)$.

Depending on the elasticity and the impact velocity, the liquid filament enters an elasto-capillary regime in which the elastic stress balances the capillary stress, and filament radius decreases exponentially. Therefore, the observed elasto-capillary thinning of the Worthington jet may provide for a simple and effective approach for estimating the extensional relaxation time of elastic fluids directly from droplet-pool interaction studies.

Further, we observe that the cavity adopts a trapezoidal shape during reversal and demonstrate that this shape evolution is driven purely by elasticity of the pool. A comparison of the cavity shape evolution in the water-glycerol and $De$ = 0.05 cases with identical viscosity shows that viscosity has a negligible effect on the formation of this shape. Furthermore, as the elasticity of the liquid increases, the cavity size decreases because a larger fraction of the kinetic energy of the cavity is converted into elastic energy via polymer stretching. Through numerical simulations, we show the distribution of elastic stresses within the jet and find that polymers in the jet core undergo greater stretching than those near the interface. As pinch-off approaches, the elastic stress increases sharply. Furthermore, for $De$=0.026, polymer stretching converts the cavity's kinetic energy into elastic energy, thereby retarding the cavity expansion and causing the cavity to propagate at approximately half the velocity observed in the pure water case. During cavity reversal, no vortex is observed inside the cavity in the elastic fluid, unlike in the pure water case, owing to the lower retraction velocity. Analysis of the local velocity field further reveals that the fluid velocity within the jet is lower than that in the pure water case.The final morphology of the Worthington jet is governed by the interplay between the growth rate of the Rayleigh instability and the inverse of the relaxation time. Depending on the elasticity of the droplet and the pool, this interplay leads to the formation of a very thin thread, a thin or thick filament, or a BOAS structure.

Our study provides new information and perspectives into the dynamics of elastic fluid droplet impact on elastic fluid pools, a phenomenon relevant to several industrial applications; including spray coating, inkjet printing, manufacturing using resin and polymer vats, and additive manufacturing, etc. Overall, we believe that our research significantly advances the fundamental understanding of the interplay between inertia, capillarity, and elasticity in controlling droplet-impact dynamics on complex fluids, and offers useful insights for applications involving elastic fluid free-surfaces and flows.

**Data availability statement:** The data pertaining to this research is available in this article.

**Conflicts of interests:** The authors declare having no conflict of interests.

**Acknowledgements**: MS thanks IIT Kharagpur for the PhD scholarship. PD thanks Anusandhan National Research Foundation (ANRF) (for the ARG grant), and IIT Kharagpur (for the internal grant) for funding this research.